\documentclass{article}

\usepackage{float}

\usepackage{scalerel}

\usepackage[backend=biber,style=nature, intitle = true]{biblatex}
\usepackage[verbose=true,letterpaper]{geometry}
\AtBeginDocument{
  \newgeometry{
    textheight=9in,
    textwidth=6.5in,
    top=1in,
    headheight=14pt,
    headsep=25pt,
    footskip=30pt
  }
}

\usepackage{upgreek}
\usepackage[symbol]{footmisc}
\usepackage[utf8]{inputenc} 
\usepackage[T1]{fontenc}    
\usepackage{hyperref}       
\usepackage{url}            
\usepackage{booktabs}       
\usepackage{amsfonts}       
\usepackage{nicefrac}       
\usepackage{microtype}      
\usepackage{xcolor}
\usepackage{textcomp}
\usepackage{authblk}

\usepackage{array}
\usepackage{graphicx}
\usepackage{esvect}
\usepackage{amsmath}

\DeclareUnicodeCharacter{2212}{-}

\usepackage{setspace}
\title{Hybrid-MPET: an open-source simulation software for hybrid electrode batteries}

\author[1]{\textbf{Qiaohao Liang}}
\author[2,3,*]{\textbf{Martin Z. Bazant}}
\affil[1]{Department of Materials Science and Engineering, Massachusetts Institute of Technology, Cambridge, MA 02139 USA}
\affil[2]{Department of Chemical Engineering, Massachusetts Institute of Technology, Cambridge, MA 02139 USA}
\affil[3]{Department of Mathematics, Massachusetts Institute of Technology, Cambridge, MA 02139 USA}
\affil[*]{Corresponding author: bazant@mit.edu}

\begin{document}

\maketitle

\begin{abstract}
As the design of single-component battery electrodes has matured, the battery industry has turned to hybrid electrodes with blends of two or more active materials to enhance battery performance. Leveraging the best properties of each material while mitigating their drawbacks, multi-component hybrid electrodes open a vast new design space that could be most efficiently explored through simulations. In this article, we introduce a mathematical modeling framework and open-source battery simulation software package for Hybrid Multiphase Porous Electrode Theory (Hybrid-MPET), capable of accounting for the parallel reactions, phase transformations and multiscale heterogeneities in hybrid porous electrodes.  Hybrid-MPET models can simulate both solid solution and multiphase active materials in hybrid electrodes at intra-particle and inter-particle scales. Its modular design also allows the combination of different active materials at any capacity fraction. To illustrate the novel features of Hybrid-MPET, we present experimentally validated models of silicon-graphite (Si-Gr) anodes used in electric vehicle batteries and carbon monofluoride (CF$_{\text{x}}$) - silver vanadium oxide (SVO) cathodes used in implantable medical device batteries. The results demonstrate the potential of Hybrid-MPET models to accelerate the development of hybrid electrode batteries by providing fast predictions of their performance over a wide range of design parameters and operating protocols.

\end{abstract}

\section*{Introduction}

Over the past 30 years, lithium-ion batteries have become the ubiquitous power source for portable electronics and electric vehicles \cite{arico2005nanostructured,chu2017path,tarascon2001issues}. 
As the next-generation applications demand further enhanced battery performance, battery research has been focused on developing new cathode and anode materials to meet the growing demand for improvements in energy and power density, high-rate performance, cycle stability, environmental friendliness, and production costs \cite{andre2015future, amine2005advanced}. It is of great scientific and practical significance for batteries today to be tailored towards the particular needs of the applications, where the performance of specialized batteries is often tuned through the use of different porous cathode and anode materials \cite{pillot2017rechargeable}. However, most widely used intercalation-based single material electrodes such as the layered transition-metal oxide LiNi$_{\text{1−x−y}}$Co$_{\text{x}}$Mn$_{\text{y}}$O$_{2}$ (NMC) cathode \cite{chakraborty2020layered, whittingham2004lithium, thackeray2021layered, park2021fictitious}, LiNi$_{\text{1−x−y}}$Co$_{\text{x}}$Al$_{\text{y}}$O$_{2}$ (NCA) cathode \cite{chakraborty2020layered, xia2018designing, radin2017narrowing}, LiFePO$_{4}$ (LFP) \cite{padhi1997phospho, kang2009battery} cathode and graphite anode \cite{guo2016li, gao2021interplay, thomas2017situ} are rapidly reaching their theoretical limits and have difficulties to meet all the required performance metrics at the same time.

To overcome the inherent limitations of single-component electrodes, researchers are increasingly focusing on developing hybrid electrodes \cite{chikkannanavar2014review, johnson2008synthesis, thackeray2007li, kang2009enhancing, gallagher2011xli2mno3, lee2011alf3, albertus2009experiments, kitao2004high, gao2009eliminating, schmidt2001future,untereker2016power,gan2005dual,chen2006hybrid,mond2014cardiac, chen2020development,jin2017challenges,yoshio2006silicon,li2021diverting,schmitt2021change,ai2022composite,cabana2010beyond,wu2017conversion, heo2022amorphous}, in which multiple active electrode materials can combine their merits to achieve better battery performances. A hybrid porous electrode is most commonly comprised of a uniform physical mixture of two or more distinct types of particles, each containing different active materials that independently react with Li$^{+}$, typically through intercalation or conversion types mechanisms during battery cycling. For example, the silicon-graphite hybrid porous anode  \cite{chen2020development,jin2017challenges,yoshio2006silicon,li2021diverting,schmitt2021change,ai2022composite} introduces small amounts of high energy density silicon particles (3579  mAh/g  for  Li$_{15}$Si$_{4}$) \cite{li2007situ, liu2011situ} into the graphite electrode (372 mAh/g for LiC$_{6}$) to considerably increase its specific capacity. A hybrid porous electrode can also contain particles
that leverage multiple parallel lithiation reactions to improve electrode performance. For example, in lithium primary batteries used to power pacemakers and implantable cardioverter-defibrillators (ICDs) \cite{schmidt2001future,untereker2016power,gan2005dual,chen2006hybrid,mond2014cardiac}, the carbon monofluoride (CF$_{\text{x}}$) - silver vanadium oxide Ag$_{2}$V$_{4}$O$_{11}$  (SVO) hybrid porous cathode takes advantage of both the high rate capability of SVO and the great specific capacity of CF$_{\text{x}}$, allowing these batteries to discharge at low currents ($\sim\upmu$A) for years without replacement and provide pulse currents ($\sim$ 3A) on demand. In SVO particles, the fast reduction reaction from V$^{5+}$ can support high current pulses, while the slower Ag$^{+}$ displacement reaction improves both the energy density and the electronic conductivity of the cathode \cite{chen2006hybrid, leising1994solid, crespi1995characterization, ramasamy2006discharge}. Note that other energy storage devices have also adopted the technique of blending multiple active materials to create high-performance hybrid electrodes: hybrid nanostructured materials that combine carbon based materials with pseudocapacitive metal oxides have been used to achieve high-performance electrochemical capacitors \cite{yu2013hybrid, simon2008materials}.

The increasing popularity of hybrid porous electrodes has led to a growing demand for improved battery simulation software capable of accounting for both the parallel reactions from individual active materials, phase transformations and multiscale heterogeneities \cite{smith2017multiphase, thomas2017situ, harris2013effects, li2014current, chueh2013intercalation, bazant2022learning, agrawal2022dynamic, agrawal2021operando, ferguson2012nonequilibrium, li2014current, bai2011suppression, ferguson2014phase} in hybrid porous electrodes. However, existing open-source battery simulation software \cite{torchio2016lionsimba,sulzer2021python,berliner2021methods,albertus2007introduction} based on Porous Electrode Theory (PET) framework \cite{newman1975porous, newman1962theoretical, doyle1993modeling, fuller1994simulation, doyle1996comparison, smith2017multiphase} cannot meet these requirements without significant modification of their software code by the users. In this article, we introduce a mathematical modeling framework and open-source battery simulation software package for Hybrid Multiphase Porous Electrode Theory (Hybrid-MPET). Building upon volume averaging methods and microscopic electrochemical relations from PET \cite{newman1975porous, newman1962theoretical, doyle1993modeling, fuller1994simulation, doyle1996comparison}, as well as the nonequilibrium thermodynamics \cite{ferguson2012nonequilibrium,ferguson2014phase, bazant2013theory,bazant2017thermodynamic,bazant2012phase,singh2008intercalation,burch2008phase,bazant2009towards,burch2009size} in Multiphase Porous Electrode Theory (MPET) \cite{smith2017multiphase}, Hybrid-MPET models are capable of accurately simulating both solid solution and multiphase active materials in the hybrid porous electrode at intra-particle and inter-particle scales. The capacity fractions, thermodynamic, kinetic, and transport properties of different active materials can also easily be adjusted due to the modular implementation of Hybrid-MPET.  We encourage reuse and modification of Hybrid-MPET, and we hope that it can serve as open-source simulation platform to aid investigation of new hybrid electrode chemistries together with experimental validation. 

The paper is organized by the following sections. In the Methods section, we describe how Hybrid-MPET addresses parallel reactions at intra-particle and inter-particle scale in greater detail. In the Results section, we highlight the novel features of
our Hybrid-MPET through sample case studies: Hybrid-MPET models are validated against experimental data from the silicon-graphite hybrid porous anode, SVO, and CF$_{\text{x}}$ - SVO hybrid porous cathode. We conclude with a summary and an outlook for future research developments in the Conclusion section.

\section*{Methods}

The main goal of the Hybrid-MPET is to create an open-source battery simulation software capable of accurately simulating the performance of hybrid electrode batteries. Hybrid-MPET falls into the category of electrochemical pseudo-two-dimensional (P2D) model, and thus utilizes the same volume averaging methods and preserves the fundamental microscopic electrochemical relations from Porous Electrode Theory (PET) \cite{newman1975porous, newman1962theoretical, doyle1993modeling, fuller1994simulation, doyle1996comparison}, including conservation of mass and charge in solid phase, liquid phases, and across interfaces. Among the existing open-source battery simulation software \cite{torchio2016lionsimba,sulzer2021python,berliner2021methods,albertus2007introduction, smith2017multiphase}, we build Hybrid-MPET by adapting Multiphase Porous Electrode Theory (MPET) \cite{smith2017multiphase} for hybrid porous electrode simulations. 

We chose to heavily reference MPET because it can accommodate both classical PET models for solid solution active materials and models for multiphase active materials based on nonequilibrium thermodynamics \cite{ferguson2012nonequilibrium,ferguson2014phase, bazant2013theory,bazant2017thermodynamic,bazant2012phase,singh2008intercalation,burch2008phase,bazant2009towards,burch2009size} in the same framework. Many commercially successful hybrid porous electrodes often contain both phase separating and solid solution materials, and we seek to select the most accurate and suitable method to describe the thermodynamics of each active material from the following two options. We could take the classical PET approach and describe thermodynamics of active materials by fitting the open circuit voltage (OCV) as a function of state of charge, which will be only accurate for solid solution active materials \cite{ferguson2012nonequilibrium,ferguson2014phase, bazant2013theory,bazant2017thermodynamic,bazant2012phase, smith2017multiphase, van2013understanding} with Fickian diffusion. For phase separating active materials such as LFP \cite{padhi1997phospho, kang2009battery} and graphite \cite{guo2016li, gao2021interplay, thomas2017situ}, often characterized by voltage plateaus in their OCV, they have multiple stable phases of different equilibrium concentrations, and such complex thermodynamic behavior cannot be described by PET models despite additional empirical modifications \cite{srinivasan2004discharge, hess2013shrinking}. Instead, for these materials, MPET and Hybrid-MPET treats their OCVs as emergent properties of multiphase materials that reflects phase separation. We can thus describe the thermodynamics of phase separating materials through phase field \cite{cahn1958free, cahn1961spinodal} models adapted for electrochemical systems \cite{ferguson2012nonequilibrium,ferguson2014phase, bazant2013theory,bazant2017thermodynamic,bazant2012phase, cogswell2012coherency,cogswell2013theory,bai2011suppression,singh2008intercalation}, where the OCV is derived from the free energy function and diffusional chemical potential  \cite{van2013understanding} of inserted Li$^{+}$ in the active material. Free energy is a function of concentration and gradients of concentration, where common tangent lines can be constructed under to represent pathway for the material to lower its energy by phase separating. Taking into account nonequilibrium
thermodynamics, MPET has thus been able to predict reaction heterogeneities in multiphase porous electrodes \cite{harris2013effects,li2014current,ferguson2012nonequilibrium,thomas2017situ,bazant2017thermodynamic, chueh2013intercalation, li2018fluid}. Such heterogeneities are experimentally observed to be highly dependent on applied current \cite{li2014current, li2018fluid, ferguson2012nonequilibrium, bai2011suppression} yet often lost over intermediate length scales as a result of volume averaging and could lead to simulation inaccuracies \cite{smith2017multiphase}. By referencing MPET, Hybrid-MPET is expected to also capture evolution of reaction heterogeneity across simulated particles in hybrid porous electrodes, which is further complicated by the existence of parallel reactions. Hybrid-MPET thus distinguishes itself from MPET \cite{smith2017multiphase} and other existing open-source battery simulation software \cite{torchio2016lionsimba,sulzer2021python,berliner2021methods,albertus2007introduction} by its ability to simulate hybrid porous electrodes with multiple active materials. We briefly revisit how reactions at particle scale are connected to macroscopic current in MPET for single material porous electrodes, and show how Hybrid-MPET introduces new equations to account for parallel reactions from different active electrode materials at intra-particle and inter-particle scale. 

In MPET, each of the $N$ finite volumes across the thickness $L$ of the porous electrode hosts $P$ active particles. For particle $p$ in finite volume $n$, $\overline{c}_{n, p} \in [0, 1]$ is used to describe its average state of charge (SOC). For intercalation reactions, SOC is known as filling fraction or a dimensionless Li$^{+}$ concentration in solid phase hosts; for primary batteries, SOC is often interchangeably used with depth of discharge or battery utilization. For non-homogeneous particles, $c$ is tracked at each discretization layer depth in the particle \cite{smith2017multiphase} to observe the effects of solid diffusion, which can be used together to reconstructed $\overline{c}$ for the whole particle. For particle $p$ in finite volume $n$, its reaction takes place independently from that of other particles, and is quantified by average volumetric reaction rate $\frac{ \partial  \overline{c}_{n, p}}{\partial t} $. 
The sum of reactions from all particles in finite volume $n$ yields its total volume averaged reaction rate $R^{V}_{n}$. The integral of the net charge consumed by the reactions across the finite volumes in the electrode will yield the macroscopic current density $i_{cell}$. Reproduced from Eq. 58-60 from MPET \cite{smith2017multiphase} and simplified by assuming Li$^{+}$ as the only cation species in lithium-ion batteries, $\frac{ \partial  \overline{c}_{n, p}}{\partial t}$ , $R^{V}_{n}$ , and $i_{cell} $ (for both anode and cathode) are defined in  Eq. \ref{eq:1} - \ref{eq:3}, respectively:
\begin{eqnarray}\label{eq:1}
\frac{ \partial  \overline{c}_{n, p}}{\partial t} = \frac{1}{V_{n, p}} \int_{A_{n, p}} j_{n, p} \,dA
\end{eqnarray}
\begin{eqnarray}\label{eq:2}
R^{V}_{n} = -(1-\epsilon)P_{L}\sum_{p = 1}^{P} \widetilde{V}_{n,p}\frac{ \partial  \overline{c}_{{n,p}}}{\partial t}  \qquad \qquad p=1,2,...,P
\end{eqnarray} 
\begin{eqnarray}\label{eq:3}
i_{cell} = \sum_{n_{a} = 1}^{N_{a}}eR^{V}_{n_{a}}\frac{L_{a}}{N_{a}} = -\sum_{n_{c} = 1}^{N_{c}}eR^{V}_{n_{c}}\frac{L_{c}}{N_{c}} \qquad \qquad n=1,2,...,N
\end{eqnarray}
where $j_{n,p}$ is the reaction flux from particle $n$ in finite volume $p$ and is equal to the net reduction current density $i_{n, p}$ scaled by (de)lithiation reaction stoichiometry. $\epsilon$ is the porosity of the electrode, $P_{L}$ is the loading percent of active material in solid phase of the electrode, which are both electrode scale parameters. $\widetilde{V}_{n, p} = \frac{V_{n, p}}{\sum_{p = 1}^{P}V_{n, p}} $ is the fraction between volume of particle $p$ and total volume of particles in finite volume $n$. 

\begin{figure}[!h]
  \centering
  \includegraphics[width = 13cm]{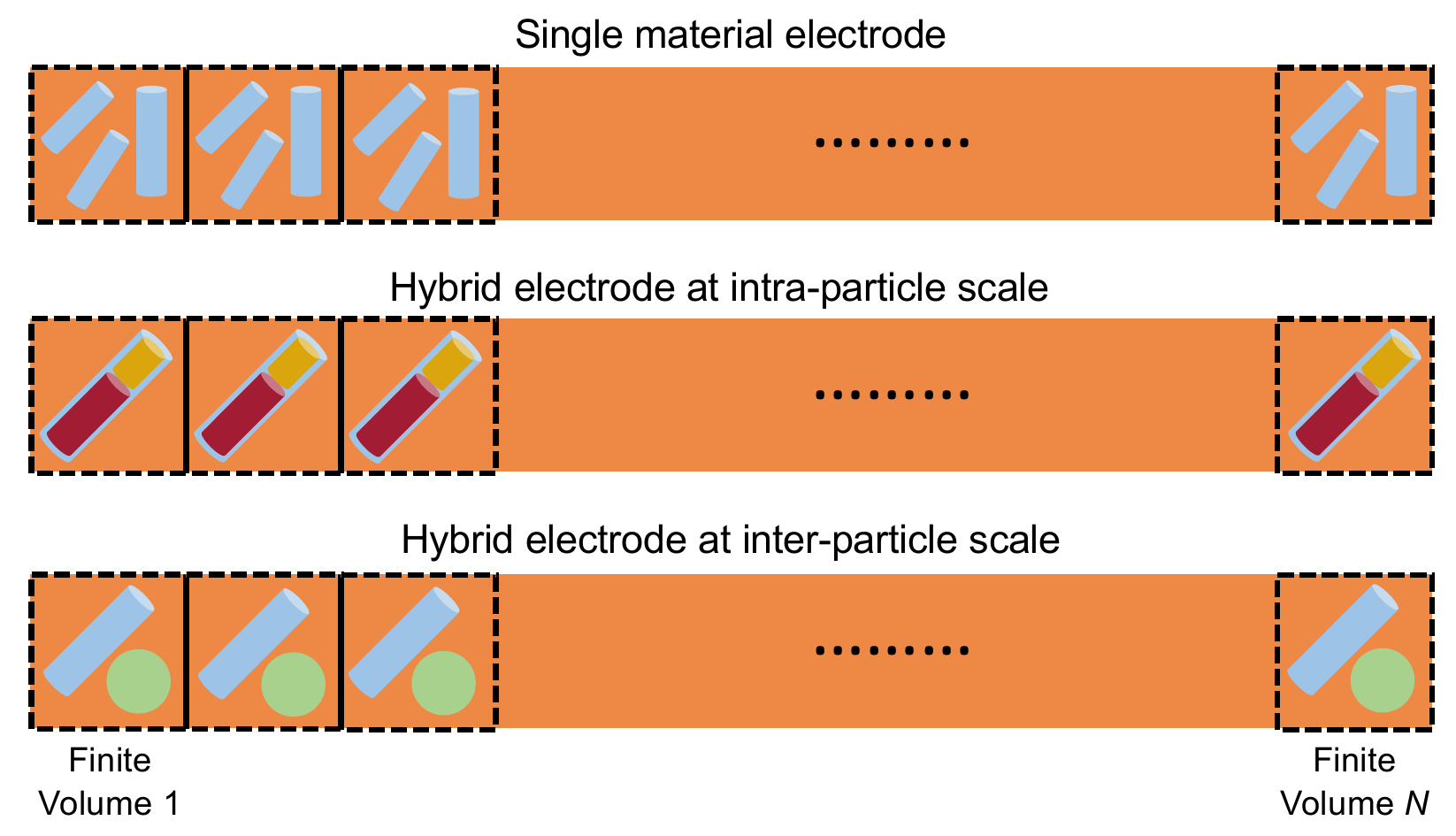}
  \caption{Schematic drawing of finite volume discretization of porous electrodes across their thickness. For single material electrodes, all particles contain the same active material. For a hybrid electrode, at intra-particle scale, each of its particles may have multiple active components; at inter-particle scale, each of its finite volume may have different particle types.}
  \label{fig:Hybridtrode}
\end{figure}

In Fig. \ref{fig:Hybridtrode}, we show a comparison between single-component electrodes and hybrid electrodes from a model formulation perspective. The complexities introduced by multiple active components and particle types mostly occur at the particle scale, and thus modifications needed for parallel reactions at intra-particle and inter-particle scale take place in Eq. \ref{eq:1} - \ref{eq:2}. The total reaction rate $R^{V}_{n}$ will be reconstructed when there are multiple active components and particle types in volume $n$. In greater detail, at intra-particle scale, we will define average state of charge $\overline{c}_{n, p}$ for a "hybrid" particle with multiple active components; at inter-particle scale, we also introduce volume correction terms to simulate hybrid electrodes made up of different active particles with any given capacity fraction. Eq. \ref{eq:3} as well as other conservation of charge and mass equations in MPET are kept unchanged for both cases below.

\subsection*{Intra-particle scale parallel reactions}
In Hybrid-MPET, simulation of intra-particle parallel reactions is crucial when modeling a hybrid electrode made up of single-type particles that allow multiple reactions. The overall reaction rate in such a "hybrid" particle takes contributions from all active components, which coexist inside the particle's volume. Within particle $p$ in finite volume $n$, $M$ different active components coexist and share the same the particle volume $V_{n, p}$ and active reaction area $A_{n,p}$. The active components have theoretical volume capacity $\rho_{1}, \rho_{2}..., \rho_{M}$, and their normalized fractions are constrained by $\sum_{m = 1}^{M}\widetilde{\rho}_{m} = 1$. Within particle $p$ in finite volume $n$, we track the state of charge of each active component separately $\overline{c}_{{n, p, 1}}$, $\overline{c}_{{n, p, 2}}$,...,$\overline{c}_{{n, p, M}}$  and thus 
\begin{eqnarray}\label{eq:4}
\overline{c}_{n,p} =\sum_{m = 1}^{M}\widetilde{\rho}_{m}\overline{c}_{n, p, m} \qquad \qquad m=1,2,...,M
\end{eqnarray}
\begin{eqnarray}\label{eq:5}
\overline{c}_{n} =\frac{\sum_{p = 1}^{P}V_{n, p}\overline{c}_{n,p}}{\sum_{p = 1}^{P}V_{n, p}}
\end{eqnarray}
Since the electrode was discretized into finite volumes of equal size, the macroscopic electrode average state of charge will just be the average of $\overline{c}_{1}$, $\overline{c}_{2}$,...,$\overline{c}_{N}$, which is equivalent to Eq. 63 in MPET \cite{smith2017multiphase}. Building upon Eq. \ref{eq:4}, the volumetric reaction rate from Eq. \ref{eq:1} is modified to,
\begin{eqnarray}\label{eq:6}
\frac{ \partial  \overline{c}_{n, p}}{\partial t} = \frac{ \partial (\sum_{m = 1}^{M}\widetilde{\rho}_{m}\overline{c}_{n, p, m})}{\partial t}  = \sum_{m = 1}^{M} \left(\widetilde{\rho}_{m} \frac{ \partial \overline{c}_{n, p, m}}{\partial t}\right) = \frac{1}{V_{n, p}}\sum_{m = 1}^{M} \left(\widetilde{\rho}_{m}  \int_{A_{n, p}} j_{n, p, m} \,dA\right)
\end{eqnarray}
where the capacity weighted sum of parallel reactions rates from all $M$ active components yields the average volumetric reaction rate. The reactions are typically described by Butler-Volmer (BV) kinetics \cite{newman2021electrochemical,bard2001fundamentals,dreyer2016new,heubner2015investigation, bazant2013theory, thomas2002mathematical} (assuming only single electron reactions) 
\begin{eqnarray}\label{eq:7}
i_{n, p, m} = j_{n, p, m} e = i_{m}(k_{m}, c_{n, p, m}, c^{l}_{n}) \left[\text{exp}\left(-\frac{\alpha_{m} e \eta^{\text{eff}}_{n, p, m}}{k_{\text{B}}T} \right) - \text{exp}\left(\frac{(1-\alpha_{m}) e \eta^{\text{eff}}_{n, p, m}}{k_{\text{B}}T}\right) \right]
\end{eqnarray}
\begin{eqnarray}\label{eq:8}
\eta^{\text{eff}}_{n, p, m} = \eta_{n, p, m} + i_{n, p, m}R_{film, m} = \Delta\phi^{s}_{n} - \Delta\phi_{m}^{eq}(c_{n, p, m}) + i_{n, p, m}R_{film, m}
\end{eqnarray}
where exchange current density $i_{m}$ is a function of rate constant $k_{m}$, symmetry coefficient $\alpha_{m}$, and Li$^{+}$ filling fractions $c_{n, p, m}$ in solid particles and Li$^{+}$ concentrations in $c^{l}_{n}$liquid electrolyte. The overpotential $\eta_{n, p, m}$ experienced by active component $m$ within particle $p$ in finite volume $n$ is the difference between electrode potential $\Delta\phi^{s}_{n}$ at finite volume $n$ and the equilibrium potential $\Delta\phi_{m}^{eq}(c_{n, p, m})$ (i.e. open-circuit voltage) of the active material $m$ within particle $p$, and we can get effective overpotential $\eta^{\text{eff}}_{n, p, m}$ after accounting for film resistance $R_{film, m}$. For each particle $p$ in the same finite volume $n$, its $M$ active components experience the same $\Delta\phi_{n}$ and are in contact with electrolyte with concentration $c^{l}_{n}$. For active component $m$, its reaction rate is determined by material specific $k_{m}$, $\alpha_{m}$, $\Delta\phi_{m}^{eq}$, $R_{film, m}$ and particle specific $c_{n, p, m}$. We have now effectively separated $M$ parallel reactions at the intra-particle scale, and therefore the total reaction rate in finite volume $n$ from Eq. \ref{eq:2} is modified to,
\begin{eqnarray}\label{eq:9}
R^{V}_{n} = -(1-\epsilon)P_{L}\sum_{p = 1}^{P} \widetilde{V}_{n, p}   \left(\sum_{m = 1}^{M} \widetilde{\rho}_{m} \frac{ \partial \overline{c}_{n, p, m}}{\partial t}\right)
\end{eqnarray}
where again $\widetilde{V}_{n, p} = \frac{V_{n, p}}{\sum_{p = 1}^{P}V_{n, p}} $ is volume fraction of particle $p$ in finite volume $n$. Instead of solving for $\overline{c}_{n,p}$ directly, we would just be solving for each $\overline{c}_{n, p, m}$ as well as $\Delta\phi^{s}_{n}$, $\Delta\phi^{l}_{n}$, $c^{l}_{n}$  at each finite volume $n$; $\overline{c}_{n,p}$ is later reconstructed from Eq. \ref{eq:4}.

\subsection*{Inter-particle scale parallel reactions}
In Hybrid-MPET, simulation of inter-particle parallel reactions is even more important because most hybrid electrodes are comprised of a uniform physical mixture of two or more distinct types of particles. Let there be $W \geq 2$ different types of particles in the hybrid porous electrode, which have theoretical volume capacity $\rho_{1}, \rho_{2}..., \rho_{W}$, respectively. We separate the thickness $L$ of the hybrid porous electrode into $N$ finite volumes, each of which hosts $P_{1}$ particles of material type 1, $P_{2}$ particles of material type 2, ..., and $P_{W}$ particles of material type $W$. Then the capacity fraction occupied by type $w$ particles in finite volume $n$ is,
\begin{eqnarray}\label{eq:10}
\widetilde{Q}_{w} = \widetilde{Q}_{n, w} = \frac{Q_{n, w}}{\sum_{w = 1}^{W}Q_{n, w}}= \frac{\sum_{p_{w} = 1}^{P_{w}} \rho_{w} V_{n, p_{w}} }{\sum_{w = 1}^{W}\left(\sum_{p_{w} = 1}^{P_{w}} \rho_{w} V_{n, p_{w}}\right)} \qquad \qquad w=1,2,...,W
\end{eqnarray}
The above setup is limited by the fact that $P_{1}$, $P_{2}$, ..., $P_{W}$ have to be all finite and positive integers $\{P_{w} \in \mathbb{Z} ^{+} | w = 1, 2, ..., W\}$. For type $w$ particles, once their sizes and theoretical volume capacity are set, often through experimental characterization or from literature documents, the capacity fraction of type $w$ active component is only dependent on its number of simulated particles $P_{w}$. To simulate hybrid porous electrodes with any given capacity fractions $\widetilde{Q}^{*}_{1}$, $\widetilde{Q}^{*}_{2}$, ..., $\widetilde{Q}^{*}_{W}$ in the electrode, we need to find $P_{1}$, $P_{2}$, ..., $P_{W}$ through solving the set of $w$ equations based on Eq. \ref{eq:10}, and obtain the following real number ratios,
\begin{eqnarray}\label{eq:11}
\frac{P_{w}}{P_{1}}= \lambda_{w1} \qquad \qquad \{\lambda_{w1} \in \mathbb{R}^{+} | w = 2, 3, ..., W\}
\end{eqnarray}
The most straightforward solution is to use a very large positive integer $P_{1}$ to also recover $P_{2}$, ..., $P_{W}$  as positive integers. 
Such large values for $P_{1}$, $P_{2}$, ..., $P_{W}$ way exceeds the number of particles necessary to capture evolution of multiscale heterogeneity or population dynamics \cite{zhao2019population, park2021fictitious, smith2017multiphase, ferguson2012nonequilibrium, ferguson2014phase} in porous electrodes. Most of the computational power would be wasted to repeatedly solve extremely similar equations for different particles of the same type in the same finite volume. A single simulation would take days if not weeks to finish, significantly limiting the efficiency in predicting the performance of hybrid porous electrodes.

To overcome this limitation, we introduce volume correction terms at inter-particle scale to simulate hybrid porous electrodes made up of different active particles with target capacity fractions $\widetilde{Q}^{*}_{1}$, $\widetilde{Q}^{*}_{2}$, ..., $\widetilde{Q}^{*}_{W}$ in the electrode. In the finite volume $n$, the volume correction terms $f_{n,w}$ will scale the volume of type $w$ particles so that capacity fraction $\widetilde{Q}^{corr}_{n,w}$ matches $\widetilde{Q}^{*}_{w}$,
\begin{eqnarray}\label{eq:12}
\widetilde{Q}^{*}_{w} = \widetilde{Q}^{corr}_{n, w} = \frac{Q^{corr}_{n, w}}{\sum_{w = 1}^{W}Q^{corr}_{n, w}}= \frac{\sum_{p_{w} = 1}^{P_{w}} \rho_{w} V_{n, p_{w}} f_{n,w} }{\sum_{w = 1}^{W}\left(\sum_{p_{w} = 1}^{P_{W}} \rho_{w} V_{n, p_{w}}f_{n,w}\right)}  \qquad \qquad w=1,2,...,W
\end{eqnarray}
$f_{n,1}$, $f_{n,2}$,...,$f_{n,W} \in \mathbb{R}^{+}$ are directly obtained by solving the set of $w$ equations based on Eq. \ref{eq:12}. Each finite volume having the correct capacity fraction ensures that the correct capacity fraction is preserved for the hybrid porous electrode at the macroscopic scale.
We still track the average SOC in each active particle separately, $\overline{c}_{n,p_{w}} \in [0, 1]$ , which now reflects the average SOC in type $w$ particles we would expect to get when type $w$ particles have the correct capacity fraction in finite volume $n$. As the volumes of type $w$ particles are weighted $f_{n,w}$, the average state of charge in volume $n$ becomes,
\begin{eqnarray}\label{eq:13}
\overline{c}^{corr}_{n} =\frac{\sum_{w = 1}^{W}\left(\sum_{p_{w} = 1}^{P_{w}}\rho_{w}V_{n, p_{w}}f_{n,w}\overline{c}_{n,p_{w}}\right)}{\sum_{w = 1}^{W}\left(\sum_{p_{w} = 1}^{P_{w}}\rho_{w}V_{n, p_{w}}f_{n,w}\right)}
\end{eqnarray}
In finite volume $n$, a Hybrid-MPET model is numerically still simulating the reactions of $P_{1}$ particles of type 1, $P_{2}$ particles of type 2, ..., and $P_{W}$ particles of type $W$, but we re-scale their contributions to the average state of charge and total reaction rate in finite volume through volume correction. As a result, we can simulate the performance of a hybrid porous electrode with any given capacity fraction. We can also still capture evolution of reaction heterogeneity and population dynamics for any particle type $w$, as long as we select $P_{w}$ to be large enough to observe such effects while small enough to finish the simulation in reasonable timescales. Again, in the same finite volume $n$, all $W$ different types of particles experience the same $\Delta\phi^{s}_{n}$ and are in contact with electrolyte with concentration $c^{l}_{n}$. For single electron reactions described by BV kinetics, they are dependent on material type specific $k_{w}$, $\alpha_{w}$, $\Delta\phi_{w}^{eq}(c_{n, p_{w}})$, $R_{film, w}$ and particle specific $c_{n, p_{w}}$,
\begin{eqnarray}\label{eq:14}
i_{n, p_{w}} = j_{n, p_{w}} e = i_{w}(k_{w}, c_{n, p_{w}}, c^{l}_{n}) \left[\text{exp}\left(-\frac{\alpha_{w} e \eta^{\text{eff}}_{n, p_{w}}}{k_{\text{B}}T}\right) - \text{exp}\left(\frac{(1-\alpha_{w}) e \eta^{\text{eff}}_{n, p_{w}}}{k_{\text{B}}T}\right) \right]
\end{eqnarray}
\begin{eqnarray}\label{eq:15}
\eta^{\text{eff}}_{n, p_{w}} = \eta_{n,p_{w}} + i_{n, p_{w}}R_{film, w} = \Delta\phi^{s}_{n} - \Delta\phi_{w}^{eq}(c_{n, p_{w}}) + i_{n, p_{w}}R_{film, w}
\end{eqnarray}
The volumetric reaction rate for particles in finite volume $n$ is kept the same as in Eq. \ref{eq:1} because volume correction is numerically applied in the total reaction rate in finite volume $n$, which now takes contribution from all $W$ types of particles, 
\begin{eqnarray}\label{eq:16}
R^{V, corr}_{n} = -(1-\epsilon)P_{L}\sum_{w = 1}^{W}\sum_{p_{w} = 1}^{P_{w}} \widetilde{V}^{corr}_{n, p_{w}} \frac{ \partial \overline{c}_{n,p_{w}}}{\partial t}
\end{eqnarray} 
where the corrected volume fractions are,
\begin{eqnarray}\label{eq:17}
\widetilde{V}^{corr}_{n, p_{w}} = \frac{V_{n,p_{w}}f_{n,w}}{\sum_{w = 1}^{W}\sum_{p_{w} = 1}^{P_{w}}V_{n,p_{w}}f_{n,w}}
\end{eqnarray} 
At each time step during the simulation on the side of this hybrid porous electrode, the solutions for $\overline{c}_{n,p_{w}}, \Delta\phi^{s}_{n}, \Delta\phi^{l}_{n}, c^{l}_{n}$ will reflect the impact of volume correction terms while simultaneously preserving conservation of charge and mass. 

Note that the equations for intra-particle scale and inter-particle scale reactions are consistent with each other. In most cases, a hybrid porous electrode consists of multiple type of particles, each type clearly distinguished by its unique single active component, and thus only the inter-particle scale equations Eq. \ref{eq:12} - \ref{eq:16} are needed. The most common example would be the silicon-graphite hybrid anode \cite{chen2020development,jin2017challenges,yoshio2006silicon,li2021diverting,schmitt2021change,ai2022composite}. Occasionally, some hybrid porous electrodes contain particles with multiple active components, leading to parallel reactions at the intra-particle scale. An example would be carbon monofluoride (CF$_{\text{x}}$) - silver vanadium oxide (SVO) hybrid cathode \cite{schmidt2001future,gan2005dual,chen2006hybrid,mond2014cardiac,leising1994solid, crespi1995characterization, ramasamy2006discharge}, which has both inter-particle scale parallel reactions from CF$_{\text{x}}$ and SVO particles, and intra-particle scale parallel reduction of V$^{5+}$ and Ag$^{+}$ in SVO. In this case, we would only need to replace $\overline{c}_{n,p_{w}}$ in Eq. \ref{eq:13} with its expression in Eq. \ref{eq:4} and $\frac{ \partial \overline{c}_{n,p_{w}}}{\partial t}$ in Eq. \ref{eq:16} with its expression in Eq. \ref{eq:6}. 

\section*{Results: Experimental validation of Hybrid-MPET models}

In this section, we demonstrate example uses of Hybrid-MPET and the predictive power of its models on hybrid porous electrodes batteries from established industrial products. The governing equations and simulation parameters are reported in Appendix A, B, and C.

\subsection*{Silicon-graphite hybrid porous electrode}

\begin{figure}[!h]
  \centering
  \includegraphics[width = 9cm]{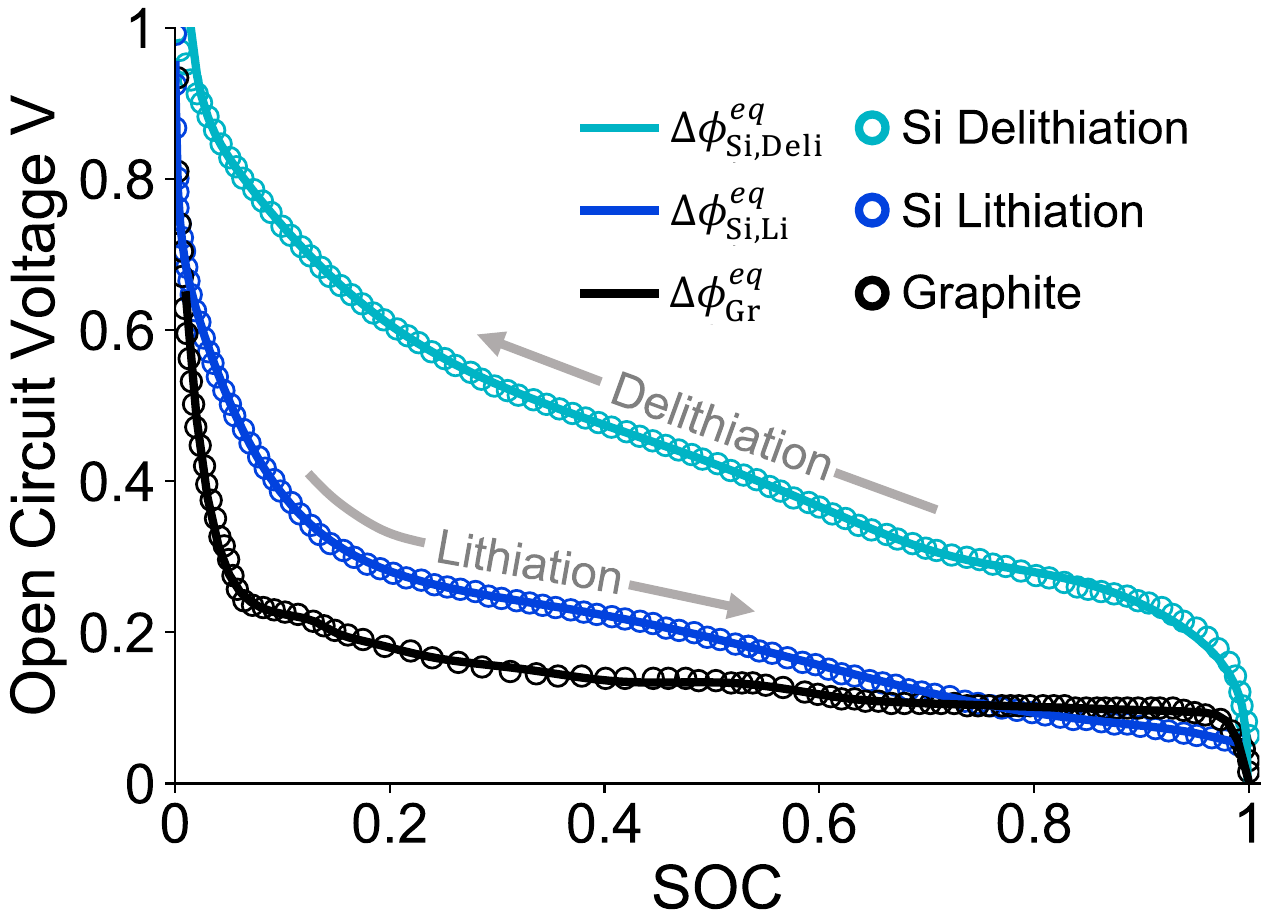}
  \caption{Open-circuit voltages as functions of state of charge for solid solution model of silicon and phase-field model of graphite. The experimentally measured OCVs of silicon\cite{verbrugge2015formulation} and   graphite \cite{, ai2020electrochemical} are also shown, respectively.}
  \label{fig:SICOCVs}
\end{figure}

A silicon-graphite (Si-Gr) hybrid porous electrode is typically used as the anode in rechargeable batteries powering electric vehicles \cite{chen2020development,jin2017challenges,yoshio2006silicon,li2021diverting,schmitt2021change,ai2022composite}. Particles of silicon or silicon oxides are uniform mixed with graphite particles to create a hybrid electrode with good cycle stability from graphite and improved energy density from introduction of silicon. Both silicon (Li$_{15}$Si$_4$) and graphite (LiC$_6$) take advantage of intercalaction reactions and undergo multiple steps of phase separation. Graphite can form many stable phases, as evident in the multiple voltage plateaus in its OCV with increased lithiation \cite{kganyago2003structural, ferguson2014phase, thomas2017situ, ohzuku1993formation, harris2010direct}. Silicon undergoes even more complex phase transformation behaviour during Li$^{+}$ de(lithiation), which are accompanied by transition between crystalline and amorphous phases, and significant volume change \cite{ai2022composite, jo2010si, muller2018quantification}. Depending on operation history, the phase and volume change of silicon lead to different stress states and give rise to its characteristic OCV hysteresis \cite{verbrugge2015formulation, lu2016voltage, wang2016investigation} as seen in Fig. \ref{fig:SICOCVs}. We recognize that complex phase-field models coupled with elasto-plastic deformation have been developed for silicon and graphite \cite{chen2014phase, cogswell2013theory}, but integrating them into the Hybrid-MPET models is beyond the scope this work. As a result, the following assumptions are used for simplicity purposes: (a) the fraction of silicon in the Si-Gr electrode is small enough such that the volume changes of silicon particles have negligible impact on electrode parameters and performance \cite{son2015silicon, liu2014pomegranate, hu2008superior, gu2012situ}. (b) silicon stays amorphous and is treated as a solid solution material with fast diffusion. (c) under the relatively low current rates below, other ohmic and transport resistances are negligible. In the future, we encourage the use of Hybrid-MPET to study interaction between the silicon and graphite at different current rates and include more accurate chemo-mechanical models to capture the multiphase behavior and stress states in silicon.

We first benchmark a Hybrid-MPET model of a Si-Gr electrode against experimental data at room temperature. The Si-Gr electrode was extracted \cite{chen2020development} from the anode of the commercial LG M50 21700 cell, and tested in a half-cell setup. In our simulation, we use an estimated 1$\%$ silicon mass fraction \cite{popp2019benchmark, popp2020ante}, which is equivalent to silicon accounting for 8.6$\%$ of the capacity fraction. We simulate a half-cell with Si-Gr electrode as the cathode, and a lithium foil counter-electrode with fast kinetics as the anode, with a generic separator in between. For reaction kinetics, classical BV kinetics for intercalation reactions are used for both silicon and graphite, whose exchange-current densities vary with both electrolyte and solid concentrations. For electrolyte transport, we use a concentrated Stefan Maxwell electrolyte model to represent a typical 1M LiPF$_{6}$ carbonate based electrolyte in lithium-ion batteries. 
For thermodynamics, we adapt the simplified one-parameter Cahn-Hillard phase field model \cite{thomas2017situ, smith2017intercalation} to capture the multiphase behavior of graphite. The formulated free energy model with three local minima \cite{smith2017intercalation} yields an open-circuit voltage that matches the sloping voltage at low graphite state of charge as well as the stage transitions of voltage at higher graphite state of charge. We resolve silicon's voltage hysteresis by using two separate OCVs for lithiation and delithiation \cite{ai2022composite} of amorphous silicon and treat it as a solid solution material. The corresponding OCVs $\Delta\phi^{eq}_{\text{Gr}}$ ,$\Delta\phi^{eq}_{\text{Si,Li}}$, $\Delta\phi^{eq}_{\text{Si,Deli}}$ can be seen in Fig. \ref{fig:SICOCVs}. We discretize the Si-Gr electrode into 10 finite volumes, each containing 1 spherical silicon particle and 1 spherical graphite particle. For most geometric and physical properties of the cell, electrode, and particles, we directly use the experimentally measured  \cite{chen2020development} values, and only refit reaction rate constants and Bruggeman exponent for better alignment between simulation and experimental data. For numerical stability purposes, for all graphite and silicon particles, we set the initial state of charge to 0.001 and 0.999, and voltage cutoff to 0.03V and 1V as end condition during lithiation and delithiation, respectively. The OCV of graphite, OCVs of silicon, and other inputs to Hybrid-MPET model of Si-Gr electrode are documented in Appendix A.

\begin{figure}[!h]
  \centering
  \includegraphics[width = \columnwidth]{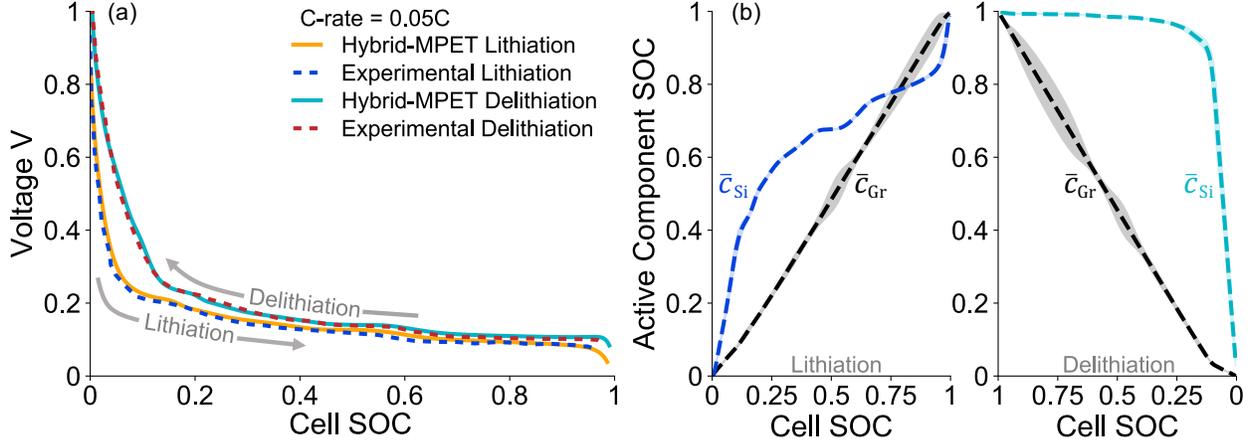}
  \caption{Hybrid-MPET simulation and experimental measurement \cite{chen2020development} of silicon-graphite hybrid electrode performance during charge and discharge at 0.05C. (a) Cell voltage vs. Cell SOC (b) Active silicon and graphite utilization vs. Cell SOC from Hybrid-MPET simulation of lithiation and delithiation. The mean SOC of silicon and graphite particles are represented in dashed lines, and are accompanied by their respective 95\% confidence intervals.}
  \label{fig:005C_arrow}
\end{figure}

In Fig. \ref{fig:005C_arrow} (a), we compare Hybrid-MPET model simulation and experimental measurement \cite{chen2020development} of Li/Si-Gr cell performance. Our models are able to match the experimental cell voltages during both discharge (lithiation) and charge (delithiation). The effect of voltage hysteresis as seen in Fig. \ref{fig:SICOCVs} can clearly be observed in the 0.05C lithiation and delithiation curves. Modelling hybrid porous electrodes introduces an additional layer of complexity through tracking the SOCs of different active species. As seen in Fig. \ref{fig:005C_arrow} (b), we show the simulated evolution of average SOC of silicon particles and graphite particles in the hybrid electrode at 0.05C operation from Hybrid-MPET model. These also match well with those predicted from other modeling and experimental studies of Si-Gr hybrid electrode at low current rates \cite{ai2022composite, lory2020probing}. While graphite shows consistent lithiation and delithiation behavior, silicon clearly shows non-linearity and asymmetry behavior, which can be directly attributed to its OCV hysteresis. During lithiation, Li$^{+}$ intercalation reactions in silicon and graphite happen in parallel, and the rates are controlled by the reaction kinetics in Eq. \ref{eq:18}. Note that in the same in finite volume $n$, silicon and graphite particles share the same $\Delta\phi^{s}_{n}$ but experience different overpotentials because silicon and graphite have different OCVs. Thus, when reaction kinetics and transport are not limited at low current rates, silicon and graphite lithiation happen in parallel throughout the discharge process, yet the observed dominant active material is determined by thermodynamics, where the one with higher OCVs typically depletes earlier. The larger the OCV difference, the more distinct dominant active component is expected to be identified. As a result, compared to that of graphite, we expect silicon SOC to increase more rapidly at the beginning of lithiation as seen in Fig. \ref{fig:005C_arrow}(b) because silicon lithiation is initially driven by a larger overpotential. Silicon and graphite SOCs will then proceed to increase individually as both particles support the total discharge current together. The cross intersection of silicon lithiation OCV and graphite OCV curves in Fig. \ref{fig:SICOCVs} suggests that silicon lithiation reaction will then be driven by smaller overpotential than that of graphite. This change results in the observed fast reaction of the residual silicon at the end of discharge process, mostly only after the full lithiation of graphite. 

The evolution of silicon and graphite SOCs during delithiation is noticeably different from those during lithiation. Because $\Delta\phi^{eq}_{\text{Gr}}$ is much lower than $\Delta\phi^{eq}_{\text{Si,Deli}}$, a clearer order of deintercalation reactions is expected: most silicon only start to delithiate when the graphite has fully delithiated, as seen in the delithiation part of Fig. \ref{fig:005C_arrow}(b). Considering silicon has faster intercalaction kinetics than graphite, we can design optimal cycling protocols for Si-Gr hybrid electrodes: for example, if we keep the half-cell setup, then during discharge of Si-Gr hybrid cathode, we should start with high rate lithiation and then switch to low rate lithiation; during charging of Si-Gr hybrid cathode, we should start with low rate delithiation and then switch to high rate delithiation. Such step-wise operation protocols ensures that we maximize both silicon and graphite utilization in hybrid electrode and minimize charging and discharging duration by taking advantage of silicon's faster kinetics. Based on the analysis above, we expect monitoring the evolution of active component SOCs in a hybrid electrode to be a valuable tool in pinpointing the dominant reactant(s) and understanding their interactions under various operational protocols. It holds promise in providing insights on how each active component contributes to the macroscopic performance and assisting optimal operation protocol design for hybrid electrodes.


\begin{figure}[!h]
  \centering
  \includegraphics[width = \columnwidth]{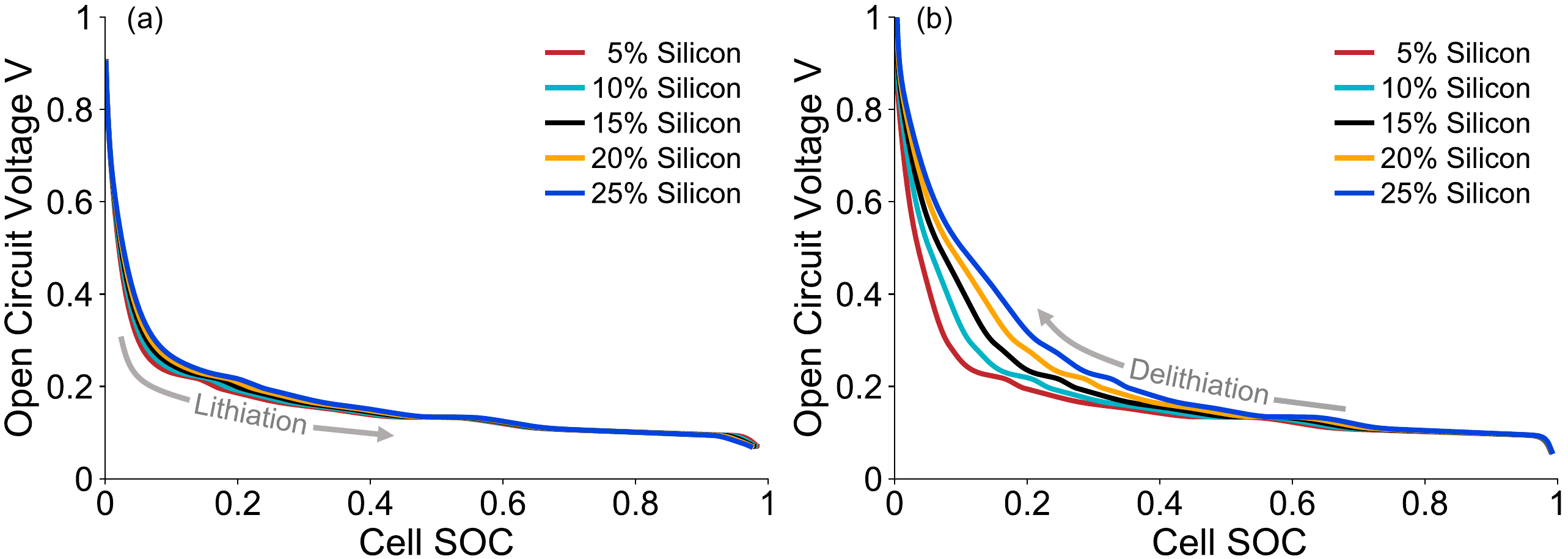}
  \caption{Hybrid-MPET prediction of the open circuit voltage of silicon-graphite hybrid electrode with varying silicon capacity fraction.}
  \label{fig:Li_Deli}
\end{figure}

Through the introduction of volume correction terms in Eq. \ref{eq:12},  Hybrid-MPET also has the versatility to predict performance of similar hybrid porous electrodes with any arbitrary targeted capacity fractions of silicon and graphite. We show the predicted OCVs of Si-Gr hybrid electrodes with different silicon capacity fraction in Fig. \ref{fig:Li_Deli}. The results were obtained by simulating lithiation and delithiation at $5\times10^{-6}$C of the same Hybrid-MPET Si-Gr hybrid electrode model as above, but now the silicon capacity fractions vary between 0.05 to 0.25. The capacity fractions and subsequently, volume fractions of silicon were kept low intentionally such that volume changes of silicon particles still have minimal impact on performance. Still, different silicon capacity fraction leads to considerable amount of change in the OCV of Si-Gr hybrid electrodes. The predicted lithiation OCVs in Fig. \ref{fig:Li_Deli}(a) resembles a mix between the $\Delta\phi^{eq}_{\text{Gr}}$ and $\Delta\phi^{eq}_{\text{Si,Li}}$. 
For delithiation in Fig. \ref{fig:Li_Deli}(b), with increased silicon capacity fraction, the characteristic voltage plateaus of graphite occupy smaller range of cell SOC and the Si-Gr delithiation OCV gradually exhibits more traits of $\Delta\phi^{eq}_{\text{Si,Deli}}$ in Fig. \ref{fig:SICOCVs}. Since Hybrid-MPET models are able to accurately predict the impact of different capacity fraction of active components on the macroscopic voltage, we believe that Hybrid-MPET can support future development of hybrid electrodes with targeted voltage profiles by providing fast predictions of performance over wide a range of design parameters, including but not limited to selection of active materials and electrolytes, capacity fractions of active components, particle sizes, electrode porosity and thickness.


\subsection*{SVO and CF$_{\text{x}}$-SVO hybrid porous electrodes}

High power density silver vanadium oxide Ag$_{2}$V$_{4}$O$_{11}$ (SVO) and high energy density carbon monofluoride (CF$_{\text{x}}$) have been traditionally used as cathode materials in lithium primary batteries powering pacemakers and ICDs, and the CF$_{\text{x}}$-SVO hybrid electrode has been developed to preserve the best properties of each \cite{schmidt2001future,gan2005dual,chen2006hybrid, untereker2016power,mond2014cardiac}. The SVO that we focus on in this paper has C-centered monoclinic unit cell \cite{zandbergen1994two, crespi1995characterization}, and is synthesized through solid state combination reaction of Ag$_{2}$O and V$_{2}$O$_{5}$ at high temperatures \cite{crespi1995characterization, crespi2001modeling}. During discharge, Li$^{+}$ insertion into 
SVO lattice is accompanied by two reversible parallel reactions: reduction of Ag$^{+}$ to metallic Ag and reduction of V$^{5+}$ to V$^{4+}$ \cite{crespi1995characterization, crespi2001modeling, leising1994solid, leifer2007nuclear}. Due to the size differences between Ag$^{+}$ and Li$^{+}$, increased Ag$^{+}$ displacement by Li$^{+}$ is accompanied by increased disorder and structural defects between vanadium oxide layers \cite{sauvage2010structural, takeuchi2001silver, onoda2001crystal, leising1994solid}. The OCV of Li/SVO \cite{crespi1995characterization, crespi2001modeling, gomadam2007modeling} has a voltage plateau at 3.24V covering the first $\frac{1}{3}$ cell capacity, strongly inferring coexistence of multiple stable phases throughout the Ag$^{+}$ displacement reaction. Another voltage plateau at 2.6V occupies the other $\frac{2}{3}$ cell capacity and corresponds to V$^{5+}$ reduction, and most studies attribute it to Li$^{+}$ intercalation into an single-phase host structure with high degree of disorder \cite{sauvage2010structural, crespi1995characterization, crespi2001modeling, west1995lithium}.  We recognize that further reduction of V$^{4+}$ to V$^{3+}$ has been observed in SVO when cell potential falls lower than 2.2V \cite{leifer2007nuclear, garcia1994lithium, crespi2001modeling}, but do not considered it because the experimental data \cite{gomadam2007modeling} we benchmark our models against do not reach such voltage ranges. As a result, in our models below, the capacity fraction of silver and vanadium are $\widetilde{\rho}_{\scaleto{\text{Ag}}{4.2pt}} = \frac{1}{3}$ and $\widetilde{\rho}_{\scaleto{\text{V}}{3.2pt}} =\frac{2}{3}$ in SVO. The composition Li$_{2}$V$_{4}$O$_{11}$ is complete silver depletion, and Li$_{6}$Ag$_{2}$V$_{4}$O$_{11}$ is full lithiation.

To show Hybrid-MPET's capability of capturing transient behavior, we next validate a model of Li/SVO half-cell and capture its voltage overshoot phenomena observed during an accelerated data collection protocol. Its experimental data \cite{gomadam2007modeling} is measured between near open-circuit low currents (5$\upmu$A) to medium currents (1000$\upmu$A) at 37$^\circ$C, which resemble the rates needed by pacemakers and ICDs to perform basic background monitoring. Such low current rates are equivalent to 3.6$\times$10$^{-6}$C and 7.2$\times$10$^{-4}$C, respectively, and would take many years to generate full voltage-capacity curves. To accelerate testing, multiple Li/SVO batteries were first discharged under 1000$\upmu$A to different depths of discharge, then immediately switched to the desired lower current to create discontinuous discharge curves that were pieced together to form complete curves \cite{gomadam2007modeling}. These unique accelerated data collection protocols have led to the observation of a voltage overshoot phenomenon: the voltage measured after Li/SVO switched from 1000$\upmu$A to a near open-circuit low current is noticeably higher than cell voltage measured when Li/SVO is continuous galvanostatic discharged under that same low current. A subsequent relaxation process is required for the cell voltage to drop down. The existing single-particle model of Li/SVO \cite{gomadam2007modeling} that treats SVO as a homogeneous solid solution material and has not been able to predict the voltage overshoot phenomena. 

Note that similar voltage overshoots have been observed for LFP electrodes \cite{sasaki2013memory, farkhondeh2014mesoscopic} in response to intermittent galvanostatic discharge, and were attributed to the division of the LFP particle population into two distinct Li-rich and a Li-poor composition groups \cite{delmas2008lithium, brunetti2011confirmation, chueh2013intercalation, robert2013multiscale} due to reaction heterogeneities \cite{farkhondeh2014mesoscopic} driven by the non-monotonic chemical potential of LFP. Since we only observe voltage overshoots in Li/SVO within its first $\frac{1}{3}$ depth of discharge, and since the existence of the 3.24V voltage plateau within first $\frac{1}{3}$ depth of discharge infers multiphase behavior, we hypothesize that introducing reaction heterogeneity during Ag$^{+}$ reduction could help capture the voltage overshoot phenomena in Li/SVO. 

One option is to introduce reaction heterogeneity within SVO particles, which would lead to coexistence of two immiscible oxide phases with different concentrations of Ag and Li, one Ag-rich, Li-poor phase and another Ag-poor, Li-rich phase \cite{west1995lithium, crespi1995characterization, crespi2001modeling}. Metallic silver would also be at equilibrium \cite{west1995lithium, crespi1995characterization, crespi2001modeling} with two phases in SVO. The phase transition process in SVO during Ag$^{+}$ reduction likely resembles those during Li$^{+}$ insertion in Cu$_{\text{y}}$TiS$_{2}$ \cite{jacobsen1989li} and Cu$_{3}$Mo$_{6}$S$_{8}$ \cite{mckinnon1984salting}, where Cu-poor, Li-rich and Cu-rich, Li-poor materials were found at equilibrium with metallic Cu. However, compared to the clear characterization of two phases from X-ray diffraction (XRD) of LFP \cite{takahashi2001characterization}, characterization of stable phases in lithiated SVO is obscured by its disordered structure \cite{leising1994solid,sauvage2010structural, crespi1995characterization}: with increased lithiation, most observe loss of crystalline SVO phase, generation of metallic silver and a highly disordered lithiated SVO phase that retains of the basic vanadium oxide layered structure with noticeable lattice parameter changes.  There is currently very limited literature and phase mapping characterization that describes the exact mechanism of electrochemically driven phase transition during Ag$^{+}$ reduction, corresponding strain effects and moving phase boundaries in lithiated SVO particles. As a result, we do not have enough information to accurately model phase separation effects within SVO particles, whether it be through shrinking core models \cite{srinivasan2004discharge} or Cahn-Hillard phase field models \cite{cahn1958free, cahn1961spinodal}.

The other option is to introduce reaction heterogeneity across the SVO particle population by creating a many-particle model \cite{dreyer2010thermodynamic, dreyer2011hysteresis, dreyer1982study} in Hybrid-MPET. In this case, instead of focusing on phase separation within a single-particle, we would expect reaction heterogeneity across the SVO particles to divide particle population into two distinct groups, one at Ag-rich, Li-poor composition and the other at Ag-poor, Li-rich composition. At this scale, reaction heterogeneity is most likely caused by thermodynamics \cite{li2014current}, kinetics \cite{park2021fictitious}, and particle size distribution \cite{chueh2013intercalation}. Since most voltage overshoots were observed at near open-circuit low rate discharge of Li/SVO cells, it is unlikely that reaction heterogeneity is caused by kinetics or concentration polarization in electrolyte when reaction across all particles are already really slow. Another consequence of discharge at such low current rates is that reaction becomes much slower than Li$^{+}$ diffusion in SVO, and the SVO particles can be conveniently treated as homogeneous \cite{gomadam2007modeling}. For industry grade SVO electrode, we expect no large size discrepancy exist across the SVO particle population and their particle sizes fall into a relatively narrow normal distribution. 

\begin{figure}[!h]
  \centering
  \includegraphics[width = \columnwidth]{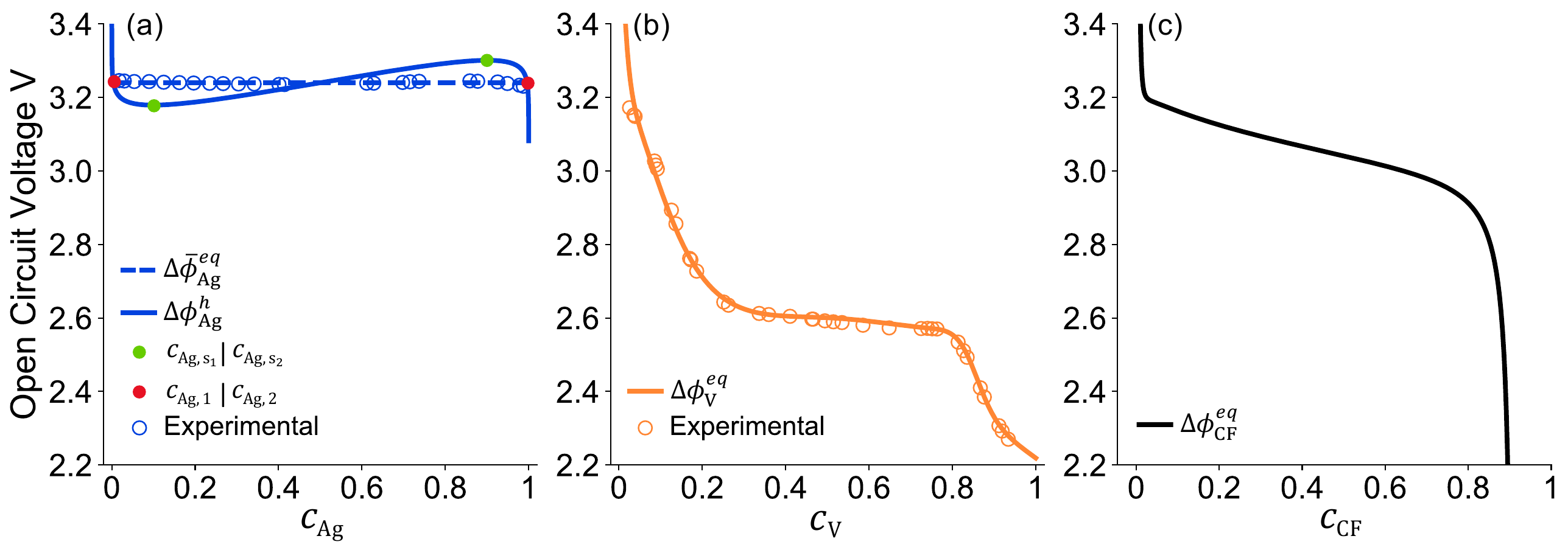}
  \caption{Open-circuit voltages as functions of state of charge for silver reduction, vanadium reduction, and carbon monofluoride reduction. The experimentally measured OCVs of SVO \cite{crespi1995characterization, crespi2001modeling, gomadam2007modeling} is split by assuming that silver and vanadium reduction are dominant in the first $\frac{1}{3}$ and last $\frac{2}{3}$ capacity of SVO, respectively. (a) The non-monotonic $\Delta\phi^{h}_{\text{Ag}}$ is derived from a double-well like free energy function, which is formulated such that a common tangent construction between its two local energy minima has a composition width that matches the width of experimentally measured voltage plateau at 3.24V during Ag$^{+}$ reduction in the first $\frac{1}{3}$ capacity of SVO. the stable compositions $c_{\scaleto{\text{Ag}}{4.5pt}, 1}$, $c_{\scaleto{\text{Ag}}{4.5pt}, 2}$ and metastable compositions $c_{\scaleto{\text{Ag}}{4.5pt}, s_{1}}$, $c_{\scaleto{\text{Ag}}{4.5pt}, s_{2}}$ are also shown.
  (b) $\Delta\phi^{eq}_{\text{V}}$ is fitted as a function of vanadium utilization from the data in last $\frac{2}{3}$ capacity of SVO. (c) $\Delta\phi^{eq}_{\text{CF}}$ is extracted from experimental data \cite{gomadam2007modeling} by assuming Tafel kinetics.}
  \label{fig:OCV_CFXSVO}
\end{figure}

We thus believe thermodynamics is the dominant driving force for creating reaction heterogeneity across SVO population at low current rates. We can create reaction heterogeneity across an ensemble of homogeneous SVO particles by introducing two energetically favorable compositions during Ag$^{+}$ reduction. To describe such thermodynamics, we formulate a double-well like homogeneous free energy as a function of only Ag state of charge $c_{\scaleto{\text{Ag}}{4.2pt}}$. The free energy has two local minima, which correspond to the stable Ag-rich, Li-poor composition $c_{\scaleto{\text{Ag}}{4.5pt}, 1}$ and Ag-poor, Li-rich composition $c_{\scaleto{\text{Ag}}{4.5pt}, 2}$. Similar to those in the many-particle models by Dreyer et al. \cite{dreyer2010thermodynamic, dreyer2011hysteresis}, the SVO particles in Hybrid-MPET model can also exchange Li$^{+}$ thought interparticle pathways to lower their total energy as an ensemble. As a result, when the ensemble composition of SVO particles arrives between $c_{\scaleto{\text{Ag}}{4.5pt}, 1}$ and $c_{\scaleto{\text{Ag}}{4.5pt}, 2}$ and the electrode is relaxed, the SVO particles tend to lower their total free energy by separating into groups with composition $c_{\scaleto{\text{Ag}}{4.5pt}, 1}$ and $c_{\scaleto{\text{Ag}}{4.5pt}, 2}$. The second derivative of free energy also identifies the two metastable compositions\cite{van2013understanding, bazant2013theory, bazant2022learning, john1961spinodal}: the Ag-rich, Li-poor one with composition $c_{\scaleto{\text{Ag}}{4.5pt}, s_{1}}$ and a Ag-poor, Li-rich one with composition $c_{\scaleto{\text{Ag}}{4.5pt}, s_{2}}$, between which separation into two compositions groups becomes spontaneous. The lowered average free energy can be found by constructing a common tangent between the two local minima in the free energy, which corresponds to a voltage plateau with width $c_{\scaleto{\text{Ag}}{4.5pt}, 2} - c_{\scaleto{\text{Ag}}{4.5pt}, 1}$ in experimentally measured OCV. Therefore, while Ag$^{+}$ reduction reaction in single SVO particles has a non-monotonic homogeneous OCV $\Delta\phi^{h}_{\text{Ag}}$, it cannot be directly obtained from measuring macroscopic cell voltage; the measured voltage plateau is instead the many-particle equilibrium potential $\Delta\overline{\phi}^{eq}_{\text{Ag}}$\cite{dreyer2010thermodynamic, dreyer2011hysteresis} that represents the emergent property of this divided SVO particle population. We can subsequently obtain a non-monotonic chemical potential and OCV for Ag$^{+}$ reduction reaction in individual SVO particles as seen in Fig. \ref{fig:OCV_CFXSVO}(a). The exact derivation \cite{bazant2013theory, bazant2017thermodynamic, smith2017multiphase}  can be found in Appendix B. 


Based on the analysis above, in our Hybrid-MPET model of Li/SVO, we define separate reaction kinetics and thermodynamics models for Ag$^{+}$ and V$^{5+}$ reduction. For reaction kinetics, we use BV kinetics for both Ag$^{+}$ and V$^{5+}$ reduction, and their exchange-current densities are fitted separately as functions of both electrolyte and solid concentrations. Compared to the conventional Li$^{+}$ intercalation reaction, it is commonly believed that the Ag$^{+}$ displacement reaction is kinetically slower \cite{sauvage2010structural}; such difference is in turn reflected in a relative smaller rate constant for Ag$^{+}$ reduction. For electrolyte transport, we use a concentrated Stefan Maxwell electrolyte model to represent a 1.1M LiAsF$_{6}$ in PC/DME in lithium-ion batteries. For thermodynamics, we use the aforementioned non-monotonic homogeneous OCV $\Delta\phi^{h}_{\text{Ag}}$ for Ag$^{+}$ reduction reaction. We fit a separate OCV $\Delta\phi^{eq}_{\text{V}}$ as a function of V$^{5+}$ depth of discharge for V$^{5+}$ reduction and treat the lithiated vanadium oxide as a solid solution material. The corresponding OCVs $\Delta\phi^{h}_{\text{Ag}}$ , $\Delta\phi^{eq}_{\text{V}}$ can be seen in Fig. \ref{fig:OCV_CFXSVO}(a)(b). We discretize the hybrid porous electrode into 100 finite volumes, each containing 10 cylindrical homogeneous SVO particles.  For numerical stability purposes, we set the initial depth of discharge of Ag$^{+}$ and V$^{5+}$ to 0.01 in all SVO particles before discharge, and voltage cutoff to 2.2V as end condition. The free energy formulation, derivation of $\Delta\phi^{h}_{\text{Ag}}$, BV reaction kinetics, and other inputs to Hybrid-MPET model of SVO electrode are documented in Appendix B. 

Note that for our Hybrid-MPET model of Li/SVO, we additionally make the following assumptions and simplifications: (a) we assume that subsequent nucleation of metallic silver atoms and growth of silver nanoparticles \cite{ramasamy2006discharge, crespi2001modeling} do not interfere with electrochemical reactions of Ag$^{+}$ and V$^{5+}$. Investigating the kinetics of nucleation and growth through using statistical models such as Kolmogorov–Johnson–Mehl–Avrami (KJMA) theory \cite{allen2008analysis, allen2007kinetic, oyama2012kinetics} is beyond the scope of the electrochemical P2D models, and we thus do not explicitly model the above processes in Hybrid-MPET.  (b) we recognize that formation of highly conductive silver nanoparticles has been observed to considerably increase the electronic conductivity of SVO electrodes and film resistance at the Li metal anode can build up over time  \cite{crespi1995characterization, crespi2001modeling, leising1994solid, ramasamy2006discharge}, but expect such effects to be only limiting for high rate applications. (c) under the low discharge rates of Li/SVO battery below, we assume that other ohmic and transport resistances in solid and liquid phases are negligible.

\begin{figure}[!h]
  \centering
  \includegraphics[width = 9cm]{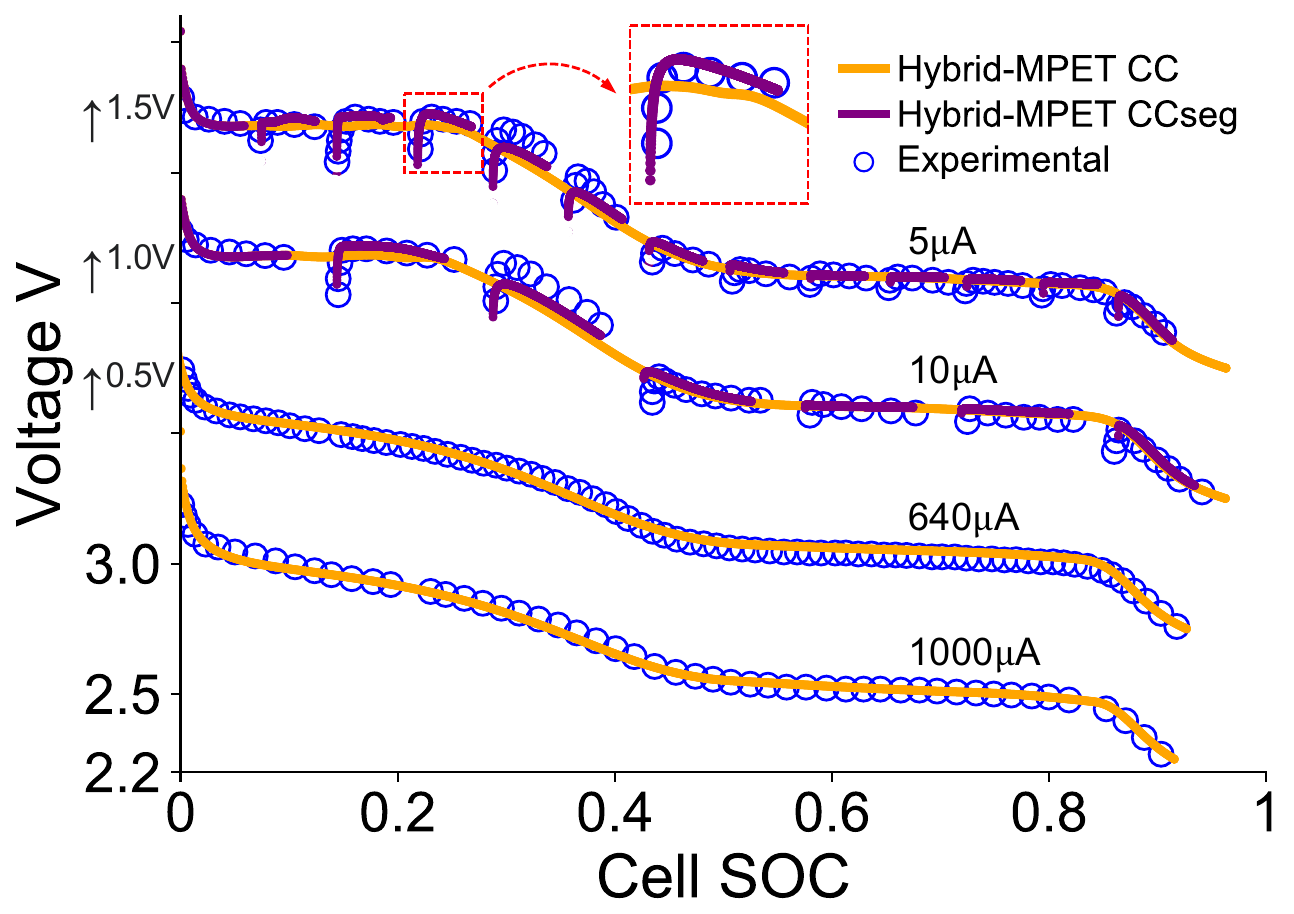}
  \caption{Hybrid-MPET model simulations and experimental measurements \cite{gomadam2007modeling} of Li/SVO under both continuous galvanostatic discharge at 5$\upmu$A, 10$\upmu$A, 640$\upmu$A, 1000$\upmu$A and accelerated data collection protocols where capacity release at 1000$\upmu$A is followed by instantaneous switch to discharging at near open-circuit low currents 5$\upmu$A or 10$\upmu$A \cite{gomadam2007modeling}. 5$\upmu$A, 10$\upmu$A, 640$\upmu$A, 1000$\upmu$A for this Li/SVO cell are equivalent to 3.6$\times$10$^{-6}$C, 7.2$\times$10$^{-6}$C, 4.6$\times$10$^{-4}$C and 7.2$\times$10$^{-4}$C. For clarity purposes, we vertically shift up the simulations and experimental data corresponding to 5$\upmu$A, 10$\upmu$A, and 640$\upmu$A by 1.5V, 1V, and 0.5V, respectively.}
  \label{fig:SVOshoot}
\end{figure}

\begin{figure}[!h]
  \centering
  \includegraphics[width = \columnwidth]{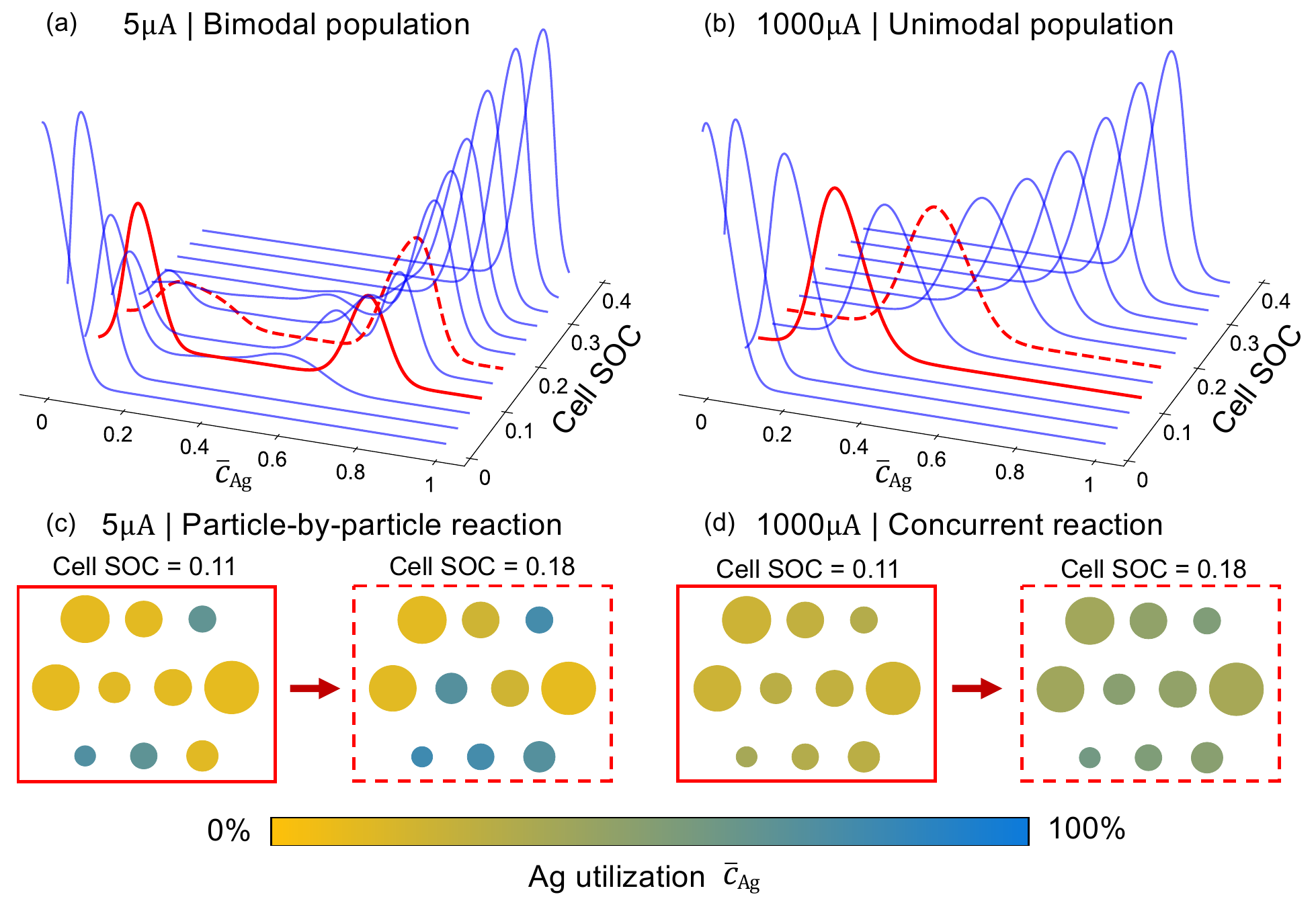}
  \caption{(a)(b) Evolution of probability density estimate of silver utilization across the SVO particle population when Li/SVO cell is discharged under constant current 5$\upmu$A and 1000$\upmu$A. Cell SOC acts as an proxy for time, and thus we can infer population dynamics by examining the gaussian kernel density estimate at different cell SOC. The firm and dashed red curve represent silver utilization distribution across the the particle ensemble at SOC = 0.11 and 0.18, respectively. (c)(d) The development of silver utilization across 10 homogeneous SVO particles in a representative finite volume from Hybrid-MPET model when discharge occurs under 5$\upmu$A and 1000$\upmu$A, respectively. For visualization clarity, we show the cylindrical SVO particles along the axial direction.}
  \label{fig:SVOpopu}
\end{figure}

In Fig. \ref{fig:SVOshoot}, we compare Hybrid-MPET model simulations and experimental measurements of Li/SVO battery discharge. We see that our Hybrid-MPET model of Li/SVO can not only accurately predict its continuous galvanostatic discharge at medium currents 640$\upmu$A, 1000$\upmu$A, but also captures the voltage overshoot phemomena prominent in the first $\frac{1}{3}$ cell capacity during accelerated data collection protocols: when the switch from 1000$\upmu$A to 5$\upmu$A or 10$\upmu$A happens, our model predicts that the cell voltage raises rapidly to a value higher than predicted the cell voltage at the same cell SOC when Li/SVO is continuous galvanostatic discharged under 5$\upmu$A or 10$\upmu$A. To explore the root cause of the voltage overshoot phenomena, we show the predicted evolution of silver utilization across the SVO particles when Li/SVO is discharged under near open-circuit low current rate 5$\upmu$A and medium current rate 1000$\upmu$A in Fig. \ref{fig:SVOpopu}. It is clear that under those two currents, the Hybrid-MPET model predicts that population dynamics of the SVO particles are very different besides at the beginning of discharge and near depletion of silver in most SVO particles. 

On the one hand, a bimodal population is observed under 5$\upmu$A discharge in Fig. \ref{fig:SVOpopu} (a), which shows the separation of SVO particle population into two groups towards Ag-rich, Li-poor and Ag-poor, Li-rich compositions. Such separation is driven by a strong reaction heterogeneity, which we introduced through using a non-monotonic $\Delta\phi^{h}_{\text{Ag}}$ derived from a double-well like free energy. Under 5$\upmu$A discharge, the current is near open-circuit, and the evolution of depth of discharge in SVO particles will follow the thermodynamically favored path. As a result, during initial stages of discharge, with increased lithiation, the compositions of SVO particles move homogeneously towards the Ag-rich, Li-poor compositions $c_{\scaleto{\text{Ag}}{4.5pt}, 1}$ and subsequently $c_{\scaleto{\text{Ag}}{4.5pt}, s_{1}}$. However, continuous discharge of the electrode means that Ag$^{+}$ needs to be reduced for Li$^{+}$ to be inserted, and composition changes have to happen somewhere between the SVO particles. The most thermodynamically favored pathway is for Ag$^{+}$ reduction and Li$^{+}$ insertion to take place only in a few particles in the porous electrode. These few particles undergo phase transition, where their compositions gradually move towards the Ag-poor, Li-rich metastable composition $c_{\scaleto{\text{Ag}}{4.5pt}, s_{2}}$ while rest of the particles remain mostly inactive at Ag-rich, Li-poor compositions. As discharge proceeds, once the compositions of the few particles reach $c_{\scaleto{\text{Ag}}{4.5pt}, s_{2}}$, a few previously inactive particles will initiate their transition from $c_{\scaleto{\text{Ag}}{4.5pt}, s_{1}}$ to $c_{\scaleto{\text{Ag}}{4.5pt}, s_{2}}$. As seen from Fig. \ref{fig:SVOpopu} (a), the fraction of active particles remains small throughout this "mosaic" particle-by-particle reaction process, which has been predicted and observed for other electrode materials \cite{li2014current, agrawal2022dynamic, agrawal2021operando, delmas2008lithium, chueh2013intercalation, dreyer2010thermodynamic, ferguson2014phase, brunetti2011confirmation}. In Fig. \ref{fig:SVOpopu} (c), by comparing the silver utilization across 10 homogeneous SVO particles in a representative finite volume from Hybrid-MPET model at two different cell SOC, the particle-by-particle reaction is clearly observed when discharging under 5$\upmu$A. Once all particles reach composition $c_{\scaleto{\text{Ag}}{4.5pt}, s_{2}}$, they then eventually reach full silver depletion as a homogeneous ensemble and only then do we start to see significant changes in vanadium depth of discharge in SVO particles. 


On the other one hand, an unimodal population is observed under 1000$\upmu$A discharge in Fig. \ref{fig:SVOpopu} (b). Reactions take place concurrently across SVO particle population and leads to homogeneous utilization of silver in the electrode. While separation of population into two composition groups is thermodynamically favored, larger current rates demand faster kinetics, and most if not all SVO particles need to concurrently \cite{li2014current} support the larger current. Examining both Fig. \ref{fig:SVOpopu} (b) and (d), we clearly see that the unimodal population is preserved throughout the discharge process until silver depletion, and the separation of particle population into two composition groups is suppressed by the relatively faster kinetics at 1000$\upmu$A.. While 1000$\upmu$A is 200$\times$ increase in current compared to 5$\upmu$A, 1000$\upmu$A (equivalent to 7.2$\times10^{-4}$C) is still low enough such that Ag$^{+}$ reduction still remains mostly dominant. The animated short videos of Fig. \ref{fig:SVOpopu} (a) and (b) can be found in the Supplementary Material.

Based on the analysis above, the root-cause of the voltage overshoot phenomena seen in the first $\frac{1}{3}$ capacity of Li/SVO becomes quite clear. The comparison will be between the following two cases at the same cell SOC: (1) immediate switch to low current 5$\upmu$A after fast capacity release under medium current 1000$\upmu$A in an accelerated data collection protocol (2) continuous discharge under 5$\upmu$A. Its 1000$\upmu$A discharge history means that case (1) has a significantly larger fraction of active particles due to previous concurrent reaction driven by medium current 1000$\upmu$A. Case (1) has more homogeneous reaction across the porous electrode while the current in case (2) is only supported by a small fraction of the porous electrode, resulting in fundamentally different battery resistant states from a more macroscale and empirical perspective. As a result, when the demand of a 5$\upmu$A current is the same, the overpotentials experienced by a large number of SVO particles in case (1) is smaller than those experienced by the smaller number of SVO particles in case (2), resulting in an observed voltage overshoot. For case (1), as the current is held at near open-circuit low rates for longer, thermodynamics soon becomes dominant over kinetics, and the previously homogeneous particle population start to separate towards two groups at Ag-rich, Li-poor and Ag-poor, Li-rich compositions despite continuous slow discharge. The previously concurrent reaction across all particles gradually transitions to the particle-by-particle like reaction as more particles reach their energetically favorable compositions, and thus the fraction of active particles gradually decreases. Such transition is predicted to be slow, and corresponds to the "relaxation" process where the overshooted voltage in case (1) to slowly returns to the cell voltage values of case (2). The animated short video on evolution of silver utilization across SVO particle population during this transition can be found in the Supplementary Material. 

While the existing single-particle model of Li/SVO \cite{gomadam2007modeling} could predict battery performance under constant current discharge, it was unable to capture the voltage overshoot phenomena because it extrapolated battery performance directly from single-particle dynamics by assuming uniform current across its electrode. In batteries, current rate dependent reaction heterogeneities \cite{bai2011suppression, li2014current} are commonly observed and often need to be introduced into battery models at particle or electrode scale to capture the correct macroscopic experimental phenomena. Because SVO was assumed to not be diffusion limited at such low current rates, only inter-particle scale reaction heterogenieties were observed and intra-particle scale concentration gradients were not considered in Hybrid-MPET model above. However, such assumptions may not hold at much high current rates. By leveraging MPET's ability to predict reaction heterogeneities at multiple scales, Hybrid-MPET has the potential to accurately predict the performance of hybrid porous electrodes over a wider range of current rates.


We finally test our Hybrid-MPET model of Li/CF$_{\text{x}}$-SVO half-cells. CF$_{\text{x}}$ is synthesized through direct fluorination of carbon \cite{zhang2015progress, liu2019brief}. In these Li/CF$_{\text{x}}$-SVO half-cells\cite{gomadam2007modeling}, it is estimated that x$\approx$1 in CF$_{\text{x}}$, and we assume CF$_{\text{x}}$ particles consist of a single CF phase. Its electrochemical reduction reaction to LiF and C has been experimentally observed to be irreversible \cite{zhang2015progress} and we thus adopt Tafel kinetics for electrochemical reduction of CF$_{\text{x}}$ \cite{gomadam2007modeling, tiedemann1974electrochemical, davis2007simulation}.  For electrolyte transport, we use a concentrated Stefan Maxwell electrolyte model to represent a 1.1M LiBF$_{6}$ in GBL/DME in lithium-ion batteries. The rate constants in BV reaction kinetics of both Ag$^{+}$ and V$^{5+}$ reduction are thus refit due to the use of a different electrolyte. For thermodynamics, the OCV of CF$_{\text{x}}$ $\Delta\phi^{eq}_{\text{CF}}$ as seen in Fig. \ref{fig:OCV_CFXSVO}(c) is extracted from experimental data \cite{gomadam2007modeling} by assuming Tafel kinetics, which is then fitted directly as a function of CF$_{\text{x}}$ utilization and resembles the CF$_{\text{x}}$ cathode's measured OCV between 3.2V and 3.4V \cite{zhang2015progress, davis2007simulation} in most non-aqueous liquid electrolytes. We discretize the hybrid porous electrode into 100 finite volumes, each containing 10 cylindrical SVO and 2 spherical CF$_{\text{x}}$ particles. For most geometric and physical properties of the cell, electrode, and particles, we reference the values from Gomadam et al \cite{gomadam2007modeling}. For numerical stability purposes, we set the initial depth of discharge of Ag$^{+}$, V$^{5+}$, and CF to 0.01 in all SVO and CF$_{\text{x}}$ particles before discharge, and voltage cutoff to 2.2V as end condition. The $\Delta\phi^{eq}_{\text{CF}}$, Tafel kinetics of CF$_{\text{x}}$, and other inputs to Hybrid-MPET model of Li/CF$_{\text{x}}$-SVO batteries are documented in Appendix C. We keep our previous assumptions and simplifications in Hybrid-MPET model of Li/SVO for Hybrid-MPET models Li/CF$_{\text{x}}$-SVO batteries, and introduce an additional one: we acknowledge that as x$\approx$1 in CF$_{\text{x}}$, C-F bonding becomes increasingly covalent, leading to poor conductivity \cite{tiedemann1974electrochemical, zhang2015progress} of CF$_{\text{x}}$ cathode. It is assumed that sufficient conductivity aid is added to the CF$_{\text{x}}$-SVO hybrid porous electrode such that the ohmic resistance from solid remains negligible at the relatively low current rates below, and the higher fraction of inactive material is reflected in lower active material volume loadings $P_{L}$ of HB1-HB4 as seen in Appendix C.

\begin{figure}[!h]
  \centering
  \includegraphics[width = \columnwidth]{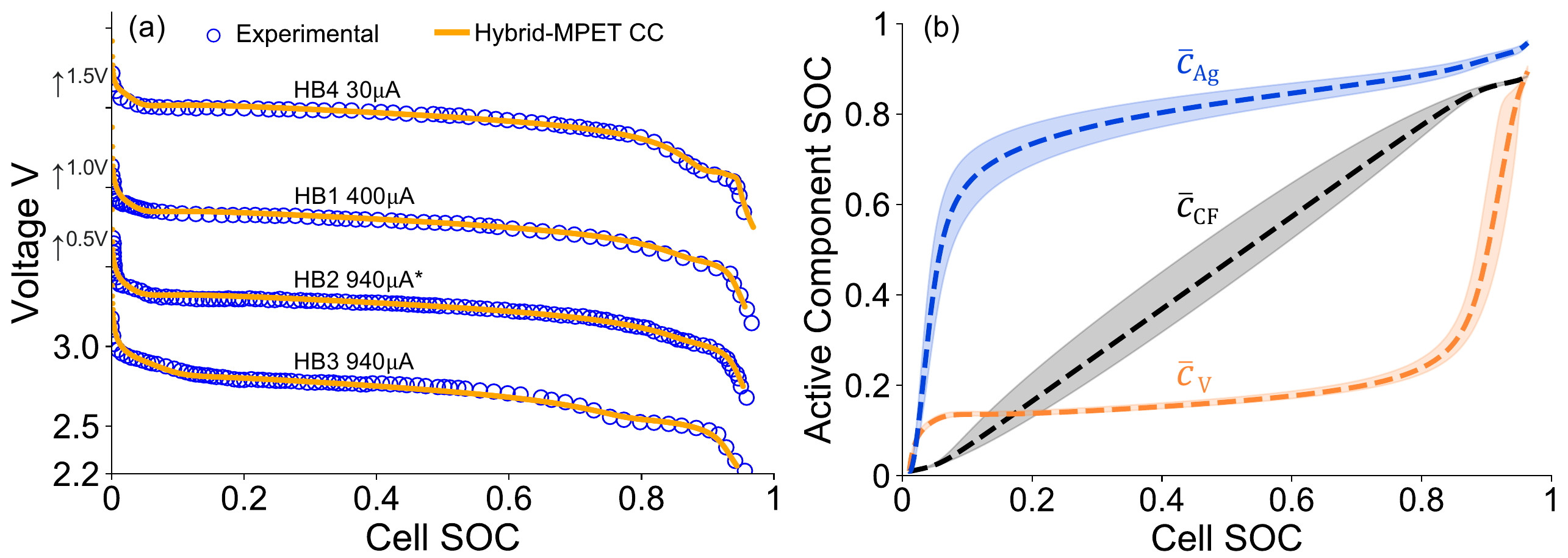}
  \caption{Hybrid-MPET prediction and experimental measurements \cite{gomadam2007modeling} of the galvanostatic discharge of hybrid cathode Li/CF$_{\text{x}}$-SVO batteries with varying CF$_{\text{x}}$ and SVO capacity fractions. (a) Cell voltage vs. Cell SOC of four Li/CF$_{\text{x}}$-SVO batteries with similar sizes under different current rates. HB1, HB2, HB4 have SVO capacity fraction 0.128, and HB3 has SVO capacity fraction 0.333. The listed discharge currents for HB1 - HB4 are equivalent to 1.9$\times10^{-5}$C, 5.2$\times10^{-4}$C, 7.9$\times10^{-4}$C, and 9.0$\times10^{-4}$C. For clarity purposes, we vertically shift up the simulations and experimental data corresponding to HB4 30$\upmu$A, HB1 400$\upmu$A, and HB2 940$\upmu$A by 1.5V, 1V, and 0.5V, respectively. (b) Active silver, vanadium, and carbon monofluoride utilization vs. Cell SOC from Hybrid-MPET simulation of HB2 Li/CF$_{\text{x}}$-SVO battery under 940$\upmu$A galvanostatic discharge. The average utilization of active silver and vanadium, and carbon monofluoride are represented in dashed lines, and are accompanied by their respective 95\% confidence intervals.}
  \label{fig:CSVO_COMP}
\end{figure}

In Fig. \ref{fig:CSVO_COMP} (a), we compare Hybrid-MPET model simulations and experimental measurements of Li/CF$_{\text{x}}$-SVO  performance. For the 4 CF$_{\text{x}}$-SVO hybrid cathodes with different geometric sizes and CF$_{\text{x}}$ and SVO capacity fractions, our model is able to accurately predict cell voltages during discharge under low to medium current rates when compared to experimental data. Compared to HB2, HB3 has considerably larger SVO capacity fraction, and such difference is accurately reflected by the Hybrid-MPET model predictions: larger SVO capacity fraction correspond to larger vanadium capacity fraction, which yields a longer characteristic voltage plateau at 2.6V for HB3. To examine the actively reacting component at different cell SOC, in Fig. \ref{fig:CSVO_COMP} (b), we show the Hybrid-MPET model predicted evolution of average depth of discharge of silver, vanadium and carbon monofluoride in the hybrid porous cathode of HB2 when discharged under 940 $\upmu$A. Again, we see that when reaction kinetics and transport are not limited at such low current rates, the three reduction reactions take place in parallel and the observed dominant active component is determined by thermodynamics: silver, vanadium, and carbon monofluoride share the same electrode potential in the same finite volume, are at equilibrium but experience different overpotentials due to each having its own OCV that is dependent on its own depth of discharge. Based on the height differences of OCVs as in Fig. \ref{fig:OCV_CFXSVO}, the expected order of depletion is silver, carbon monofluoride, and vanadium, which is indeed observed in Fig. \ref{fig:CSVO_COMP} (b). During initial stages of discharge, Ag$^{+}$ reduction is dominant while V$^{+}$ reduction takes place in parallel to maintain equilibrium potential with silver in SVO. This prediction matches observations from many experimental studies \cite{leising1994solid, crespi1995characterization, gomadam2007modeling, sauvage2010structural, ramasamy2006discharge, grisolia2011density}, which have argued that V$^{5+}$ reduction could precede Ag$^{+}$ reduction and support that they are more parallel rather than sequential. Meanwhile, no significant reaction heterogeneity is observed across SVO particles, and the separation of SVO particle population into two composition groups is again suppressed by the relatively faster kinetics at 940$\upmu$A. The loss of capacity brings the cell voltage down to values where CF$_{\text{x}}$ becomes the primary active component to support the discharge current through its reduction, during which most residual silver is depleted while vanadium utilization stays nearly constant. When CF$_{\text{x}}$ reaches near depletion, V$^{5+}$ reduction becomes dominant until the end of discharge. The Hybrid-MPET model prediction of silver, vanadium and carbon monofluoride utilization as seen in Fig. \ref{fig:CSVO_COMP} (b) match well with those from other modeling studies of Li/CF$_{\text{x}}$-SVO batteries \cite{gomadam2007modeling}. Note that the observed depletion order is expected to be true only under low current rates where thermodynamics are dominant over kinetics and transport limitations are neglected; it is highly likely that such depletion order is no longer held at higher current rates. For example, it has been experimentally observed \cite{chen2006hybrid,mond2014cardiac} that at high current pulses, SVO provides most of the power due to its higher rate capability while during post-pulse relaxation at low currents, CF$_{\text{x}}$ discharges itself to recharge SVO until both are at equilibrium. Such complicated interactions between different active materials offers great opportunities for future modeling studies using Hybrid-MPET.

\vspace{-10pt}

\section*{Conclusion}

Despite the increasing research and application interest in hybrid electrode batteries, there is a lack of open-source battery simulation software that are suited for them. To address this issue, we present Hybrid-MPET, a mathematical modeling framework and open-source battery simulation software package. Hybrid-MPET is implemented based on the MPET framework \cite{smith2017multiphase}, and its key modifications have been described in detail. Through sample case studies of Si-Gr, SVO, and CF$_{\text{x}}$-SVO hybrid porous electrodes, we have illustrated the novel features of Hybrid-MPET models, including but not limited to its ability to: (1) account for intra-particle scale and inter-particle scale parallel reactions (2) accommodate both phase separating and solid solution type active electrode materials in the same simulation (3) allow free combination of active materials of interest at any capacity fraction in the hybrid electrode. (4) predict the evolution of state of charge for each active material through charge and discharge processes (5) capture the impact of particle scale reaction heterogeneity on macroscopic battery performance. 

Since Hybrid-MPET is open-sourced and inherits the modular design from MPET, we encourage its reuse, modification, and optimization. Modelling hybrid porous electrodes is exciting but also challenging because they offer an entire new dimension of complexity, where the state of charge needs to be tracked separately for different active materials. Despite reactions from each active material being parallel, the clear sequential depletion orders observed in the sample simulations are mostly true at low current rates, where thermodynamics are dominant. However, at higher current rates, the limitations from kinetics and transport is expected to have greater impact. 

A natural extension of this work would be towards using Hybrid-MPET models to simulate performance of hybrid electrode at higher current rates; we also expect high currents to drive the its active materials to far-from-equilibrium states, and have great interest in studying capacity exchange between the different active materials during post-pulse relaxation. As hybrid electrodes open a vast new design space and require more comprehensive testing protocols, we believe Hybrid-MPET has great potential to complement experimental research and accelerate the future development of hybrid electrodes with targeted voltage profiles by providing fast and accurate predictions of battery performance.

Another important extension would be to develop degradation models for Hybrid MPET. The literature on battery degradation has focused almost exclusively on single-component electrodes, notably graphite, using coarse-grained models of the key capacity-fade mechanisms, such as solid-electrolyte-interphase (SEI) growth~\cite{pinson2012theory} and electrochemomechanical fatigue~\cite{jana2022physics}. Both of these phenomena are much stronger in silicon than in graphite, so the blending of silicon in graphite to achieve higher energy density must be balanced against the acceleration of capacity fade~\cite{dose2018capacity,dhillon2021modelling}. Hybrid cathodes involving different transition metal oxides will also undergo oxidative surface degradation at different rates, depending on their stochiometries~\cite{zhuang2022theory}.  Some critical degradation reactions, such as lithium metal plating in graphite anodes (also affecting battery safety), are also known to be strongly coupled with phase separation dynamics~\cite{harris2010direct,thomas2017situ,gao2021interplay,finegan2020spatial}. The coupling of different degradation mechanisms and rates for the various components of hybrid electrodes will have a nontrivial dependence on microstructure and composition, which could be unraveled more easily with the help of Hybrid-MPET simulations that also account for heterogeneity arising from phase transformations.  

\section*{Code availability}
The code for Hybrid-MPET battery simulation software is available in the following GitHub repository \cite{hybridMPETGithub}: https://github.com/HarryQL/Hybrid-MPET


\section*{Acknowledgements}
This work is supported by Medtronic Public Limited Company.

\section*{Appendix A}
We postulate a free energy functional describing the physics of the graphite particle, and separate it into homogeneous and non-homogenous free energy. Following van der
Waals \cite{rowlinson1979translation} and Cahn and Hilliard \cite{cahn1958free}, we use a simple gradient penalty
term to describe the non-homogeneous free energy, and obtain \cite{smith2017multiphase, thomas2017situ} the following  the diffusional chemical potential $\mu_{\scaleto{\text{Gr}}{4.2pt}}$ of Li$^{+}$ in inserted into graphite is defined as,  
\begin{eqnarray}\label{eq:90}
\mu_{\scaleto{\text{Gr}}{4.2pt}}(c_{\scaleto{\text{Gr}}{4.2pt}}) = \frac{\updelta G}{\updelta c_{\scaleto{\text{Gr}}{4.2pt}}} 
= \frac{\partial g}{\partial c_{\scaleto{\text{Gr}}{4.2pt}}} - \nabla \cdot \frac{\partial g}{\partial \nabla c_{\scaleto{\text{Gr}}{4.2pt}}}= -0.02 + \mu_{1} + \mu_{2} + \mu_{3} + \mu_{4} + \mu_{5} - \frac{\kappa}{\rho_{\scaleto{\text{Gr}}{4.2pt}}} \nabla^{2} c_{\scaleto{\text{Gr}}{4.2pt}}
\end{eqnarray}
\begin{multline}\label{eq:91}
\mu_{1} = k_{\text{B}}T \left[ \left(-30\text{exp}\left(-\frac{c_{\scaleto{\text{Gr}}{4.2pt}}}{0.025}\right) - 2(1-c_{\scaleto{\text{Gr}}{4.2pt}})\right)S_{D}(c_{\scaleto{\text{Gr}}{4.2pt}}, 0.38, 0.05)
+ 0.7\left(\text{tanh}\left(\frac{c_{\scaleto{\text{Gr}}{4.2pt}}-0.37}{0.075}\right) - 1\right) \right.\\
\left.
+ 0.8\left(\text{tanh}\left(\frac{c_{\scaleto{\text{Gr}}{4.2pt}}-0.2}{0.06}\right) - 1\right)
+ 0.38\left(\text{tanh}\left(\frac{c_{\scaleto{\text{Gr}}{4.2pt}}-0.14}{0.015}\right) - 1\right)
\right]
S_{D}(c_{\scaleto{\text{Gr}}{4.2pt}}, 0.42, 0.05)
\end{multline}
\begin{eqnarray}\label{eq:92}
\mu_{2} = -k_{\text{B}}T\frac{0.05}{c_{\scaleto{\text{Gr}}{4.2pt}}^{0.55}} 
\end{eqnarray}
\begin{eqnarray}\label{eq:93}
\mu_{3} = 10k_{\text{B}}TS_{U}(c_{\scaleto{\text{Gr}}{4.2pt}}, 1, 0.015) 
\end{eqnarray}
\begin{eqnarray}\label{eq:94}
\mu_{4} = 1.8\widetilde{\Omega}_{a}k_{\text{B}}T (0.17 - c_{\scaleto{\text{Gr}}{4.2pt}} ^ {0.98})
S_{D}(c_{\scaleto{\text{Gr}}{4.2pt}}, 0.55, 0.045) 
S_{U}(c_{\scaleto{\text{Gr}}{4.2pt}}, 0.38, 0.05) 
\end{eqnarray}
\begin{eqnarray}\label{eq:95}
\mu_{5} = k_{\text{B}}T \left[0.4\widetilde{\Omega}_{a} (0.74 - c_{\scaleto{\text{Gr}}{4.2pt}})+ 0.55\widetilde{\Omega}_{b} - 2(1-c_{\scaleto{\text{Gr}}{4.2pt}}) \right]
S_{U}(c_{\scaleto{\text{Gr}}{4.2pt}}, 0.6, 0.04) 
\end{eqnarray}
where $\widetilde{\Omega}_{a}=0.81169$, $\widetilde{\Omega}_{b}=2.2214$, $k_{\text{B}} =  1.38\times10^{-23} \frac{J}{K}$, and $T =  298.15 K$. 
Step-up and step-down functions are defined as \cite{thomas2017situ},
\begin{eqnarray}\label{eq:96}
S_{D}(c_{\scaleto{\text{Gr}}{4.2pt}}, c_{\text{x}}, \delta) = 0.5 \left(-\text{tanh} \left(\frac{c_{\scaleto{\text{Gr}}{4.2pt}} - c_{\text{x}}}{\delta}\right) + 1\right)
\end{eqnarray}
\begin{eqnarray}\label{eq:97}
S_{U}(c_{\scaleto{\text{Gr}}{4.2pt}}, c_{\text{x}}, \delta) = 0.5 \left(\text{tanh} \left(\frac{c_{\scaleto{\text{Gr}}{4.2pt}} - c_{\text{x}}}{\delta}\right) + 1\right)
\end{eqnarray}
Subsequently, the OCV for graphite is,
\begin{eqnarray}\label{eq:98}
\Delta\phi^{eq}_{\text{Gr}}(c_{\scaleto{\text{Gr}}{4.2pt}}) = 0.12 - \frac{\mu_{\scaleto{\text{Gr}}{4.2pt}}(c_{\scaleto{\text{Gr}}{4.2pt}})}{e} 
\end{eqnarray}
where $e = 1.602\times10^{-19}C$, and 0.12 volts is a fitted reference potential $\Delta\phi^{\circ}_{\text{Gr}}(c_{\scaleto{\text{Gr}}{4.2pt}})$ \cite{bazant2013theory, smith2017multiphase}.

The OCVs for silicon lithiation and delithiation are fitted separately as,
\begin{multline}\label{eq:99}
\Delta\phi^{eq}_{\text{Si,Li}} = 0.284 -0.084\text{ln}\left(\frac{c_{\scaleto{\text{Si}}{4.2pt}}}{1-c_{\scaleto{\text{Si}}{4.2pt}}}\right)  \\
+ \frac{
0.022c_{\scaleto{\text{Si}}{4.2pt}} 
-0.711c_{\scaleto{\text{Si}}{4.2pt}}^2
+ 2.673c_{\scaleto{\text{Si}}{4.2pt}}^3
-3.762 c_{\scaleto{\text{Si}}{4.2pt}}^4
+ 0.246c_{\scaleto{\text{Si}}{4.2pt}}^5
+ 3.588c_{\scaleto{\text{Si}}{4.2pt}}^6
-2.050c_{\scaleto{\text{Si}}{4.2pt}}^7}
{
0.007
+0.131c_{\scaleto{\text{Si}}{4.2pt}} 
+1.158c_{\scaleto{\text{Si}}{4.2pt}}^2
-1.120c_{\scaleto{\text{Si}}{4.2pt}}^3
-0.290 c_{\scaleto{\text{Si}}{4.2pt}}^4
+ 0.790c_{\scaleto{\text{Si}}{4.2pt}}^5
-0.657c_{\scaleto{\text{Si}}{4.2pt}}^6}
\end{multline}

\begin{multline}\label{eq:100}
\Delta\phi^{eq}_{\text{Si,Deli}} = 0.948 -0.006\text{ln}\left(\frac{c_{\scaleto{\text{Si}}{4.2pt}}}{1-c_{\scaleto{\text{Si}}{4.2pt}}}\right)  \\
+ \frac{
-1.093c_{\scaleto{\text{Si}}{4.2pt}} 
+2.886c_{\scaleto{\text{Si}}{4.2pt}}^2
-1.670c_{\scaleto{\text{Si}}{4.2pt}}^3
-2.133 c_{\scaleto{\text{Si}}{4.2pt}}^4
+ 0.529c_{\scaleto{\text{Si}}{4.2pt}}^5
+ 1.895c_{\scaleto{\text{Si}}{4.2pt}}^6
-0.509c_{\scaleto{\text{Si}}{4.2pt}}^7}
{
0.362
+0.230c_{\scaleto{\text{Si}}{4.2pt}} 
-2.027c_{\scaleto{\text{Si}}{4.2pt}}^2
+1.568c_{\scaleto{\text{Si}}{4.2pt}}^3
+1.181 c_{\scaleto{\text{Si}}{4.2pt}}^4
+ 1.046c_{\scaleto{\text{Si}}{4.2pt}}^5
-2.249c_{\scaleto{\text{Si}}{4.2pt}}^6}
\end{multline}
The reaction kinetics of graphite and silicon are,
\begin{eqnarray}\label{eq:101}
i_{\text{\scaleto{\text{Gr}}{4.2pt}}} = j_{\scaleto{\text{Gr}}{4.2pt}} e = k_{\scaleto{\text{Gr}}{4.2pt}} (c^{l})^{0.5} c_{\text{\scaleto{\text{Gr}}{4.2pt}}}^{0.5}(1-c_{\text{\scaleto{\text{Gr}}{4.2pt}}})^{0.5}
\left[\text{exp}\left(-\frac{\alpha_{\text{\scaleto{\text{Gr}}{4.2pt}}} e \eta^{\text{eff}}_{\text{\scaleto{\text{Gr}}{4.2pt}}}}{k_{\text{B}}T}\right) - \text{exp}\left(\frac{(1-\alpha_{\text{\scaleto{\text{Gr}}{4.2pt}}}) e \eta^{\text{eff}}_{\text{\scaleto{\text{Gr}}{4.2pt}}}}{k_{\text{B}}T}\right) \right]
\end{eqnarray}
\begin{eqnarray}\label{eq:102}
i_{\text{\scaleto{\text{Si}}{4.2pt}}} = j_{\scaleto{\text{Si}}{4.2pt}} e = k_{\scaleto{\text{Si}}{4.2pt}} (c^{l})^{0.5} c_{\text{\scaleto{\text{Si}}{4.2pt}}}^{0.5}(1-c_{\text{\scaleto{\text{Si}}{4.2pt}}})^{0.5}
\left[\text{exp}\left(-\frac{\alpha_{\text{\scaleto{\text{Si}}{4.2pt}}} e \eta^{\text{eff}}_{\text{\scaleto{\text{Si}}{4.2pt}}}}{k_{\text{B}}T}\right) - \text{exp}\left(\frac{(1-\alpha_{\text{\scaleto{\text{Si}}{4.2pt}}}) e \eta^{\text{eff}}_{\text{\scaleto{\text{Si}}{4.2pt}}}}{k_{\text{B}}T}\right) \right]
\end{eqnarray}
The separator used has thickness 12$\upmu$m, porosity 0.47 \cite{chen2020development}, and tortuosity $\epsilon^{-1}$, and is discretized into 2 finite volumes.The other key parameters used in Hybrid-MPET model of Li/Si-Gr can be seen in Table \ref{table:SICelectrode},
\renewcommand{\thetable}{1}
\renewcommand{\arraystretch}{1.2}
\newcolumntype{P}[1]{>{\centering\arraybackslash}p{#1}}

\begin{table}[!h]
  \caption{Geometric and materials properties used in simulating silicon-graphite hybrid porous electrode}
  \centering
  \begin{tabular}{ c c c c c}
  \hline
   Parameter &Description & Unit & \multicolumn{2}{c}{Si-Gr Hybrid Porous Electrode} 
   \\
  \hline
    $L$ & thickness & $\upmu$m &    \hspace{+80pt} 85.2 \cite{chen2020development} \\
    $\epsilon$ & porosity & - & \hspace{+80pt} 0.25 \cite{chen2020development}   
    \\
    $P_{L}$ & active solid volume fraction & - & \hspace{+80pt} 0.87 
    \\
    $\tau$ & tortuosity & - & \hspace{+80pt} $\epsilon^{-0.2}$
    \\
    $\widetilde{Q}_{Si} $ & capacity fraction of silicon & -  & \hspace{+80pt} 0.086 
    \\
    $\widetilde{Q}_{Gr} $ & capacity fraction of graphite & -  & \hspace{+80pt} 0.914 
    \\
      
  \hline
        Parameter &Description & Unit & Graphite & Silicon \\
  \hline
    $R$ & particle radius & $\upmu$m  & $\mathcal{N}$(5.86, 1.2) \cite{chen2020development} & $\mathcal{N}$(1.52, 0.8) \cite{chen2020development}
    \\
    $\rho$ & volumetric site density & mol m$^{-3}$ & 29700 \cite{ai2020electrochemical} & 277990 \cite{liu2020multiphysics}  \\
    $D$ & diffusion coefficient & m$^{2}$s$^{-1}$ &5.5$\times$10$^{-14}$ \cite{ai2022composite} & - 
    \\
    $\alpha$ & symmetry coefficient & - & 0.5 & 0.5
    \\
    $k$ & rate constant & A m$^{-2}$ & 1 & 40
    \\
    \hline
    \end{tabular}
  \label{table:SICelectrode}
\end{table}

Adopting  Eq. \ref{eq:13} and \ref{eq:16} for Si-Gr, we have the following inter-particle scale equations are,
\begin{eqnarray}\label{eq:18}
\overline{c}^{corr}_{n} =\frac{\sum_{p_{\scaleto{\text{Si}}{2.7pt}} = 1}^{P_{\scaleto{\text{Si}}{2.7pt}}}\rho_{\scaleto{\text{Si}}{3.2pt}}V_{n, p_{\scaleto{\text{Si}}{2.7pt}}}f_{n,\scaleto{\text{Si}}{3.2pt}}\overline{c}_{n,p_{\scaleto{\text{Si}}{2.7pt}}} + \sum_{p_{\scaleto{\text{Gr}}{2.7pt}} = 1}^{P_{\scaleto{\text{Gr}}{2.7pt}}}\rho_{\scaleto{\text{Gr}}{3.2pt}}V_{n, p_{\scaleto{\text{Gr}}{2.7pt}}}f_{n,\scaleto{\text{Gr}}{3.2pt}}\overline{c}_{n,p_{\scaleto{\text{Gr}}{2.7pt}}}
}{\sum_{p_{\scaleto{\text{Si}}{2.7pt}} = 1}^{P_{\scaleto{\text{Si}}{2.7pt}}}\rho_{\scaleto{\text{Si}}{3.2pt}}V_{n, p_{\scaleto{\text{Si}}{2.7pt}}}f_{n,\scaleto{\text{Si}}{3.2pt}} + \sum_{p_{\scaleto{\text{Gr}}{2.7pt}} = 1}^{P_{\scaleto{\text{Gr}}{2.7pt}}}\rho_{\scaleto{\text{Gr}}{3.2pt}}V_{n, p_{\scaleto{\text{Gr}}{2.7pt}}}f_{n,\scaleto{\text{Gr}}{3.2pt}}}
\end{eqnarray}
\begin{eqnarray}\label{eq:19}
R^{V, corr}_{n} = -(1-\epsilon)P_{L}\left(
\sum_{p_{\scaleto{\text{Si}}{3.2pt}} = 1}^{P_{\scaleto{\text{Si}}{3.2pt}}} \widetilde{V}_{n, p_{\scaleto{\text{Si}}{2.7pt}}} \frac{ \partial \overline{c}_{n,p_{\scaleto{\text{Si}}{2.7pt}}}}{\partial t}
+
\sum_{p_{\scaleto{\text{Gr}}{3.2pt}} = 1}^{P_{\scaleto{\text{Gr}}{3.2pt}}} \widetilde{V}_{n, p_{\scaleto{\text{Gr}}{2.7pt}}} \frac{ \partial \overline{c}_{n,p_{\scaleto{\text{Gr}}{2.7pt}}}}{\partial t}
\right) 
\end{eqnarray} 
where the corrected volume fractions are
\begin{eqnarray}\label{eq:20}
\widetilde{V}_{n, p_{\scaleto{\text{Si}}{2.7pt}}} = \frac{V_{n, p_{\scaleto{\text{Si}}{2.7pt}}}f_{n,\scaleto{\text{Si}}{3.2pt}}}{\sum_{p_{\scaleto{\text{Si}}{2.7pt}} = 1}^{P_{\scaleto{\text{Si}}{2.7pt}}}V_{n,p_{\scaleto{\text{Si}}{2.7pt}}}f_{n,\scaleto{\text{Si}}{3.2pt}} 
+
\sum_{p_{\scaleto{\text{Gr}}{2.7pt}} = 1}^{P_{\scaleto{\text{Gr}}{2.7pt}}}V_{n,p_{\scaleto{\text{Gr}}{2.7pt}}}f_{n,\scaleto{\text{Gr}}{3.2pt}} }
\end{eqnarray} 
\begin{eqnarray}\label{eq:21}
\widetilde{V}_{n, p_{\scaleto{\text{Gr}}{2.7pt}}} = \frac{V_{n, p_{\scaleto{\text{Gr}}{2.7pt}}}f_{n,\scaleto{\text{Gr}}{3.2pt}}}{\sum_{p_{\scaleto{\text{Si}}{2.7pt}} = 1}^{P_{\scaleto{\text{Si}}{2.7pt}}}V_{n,p_{\scaleto{\text{Si}}{2.7pt}}}f_{n,\scaleto{\text{Si}}{3.2pt}} 
+
\sum_{p_{\scaleto{\text{Gr}}{2.7pt}} = 1}^{P_{\scaleto{\text{Gr}}{2.7pt}}}V_{n,p_{\scaleto{\text{Gr}}{2.7pt}}}f_{n,\scaleto{\text{Gr}}{3.2pt}} }
\end{eqnarray}

\section*{Appendix B}

To describe the thermodynamics of Ag$^{+}$ reduction, we postulate a double-well like free energy functional without non-homogeneous terms (e.g. gradient penalty or coherent stress), 
\begin{eqnarray}\label{eq:104}
g^{h} =  k_{\text{B}}T \left[c_{\scaleto{\text{Ag}}{4.2pt}}\text{ln}(c_{\scaleto{\text{Ag}}{4.2pt}}) + (1-c_{\scaleto{\text{Ag}}{4.2pt}})\text{ln}(1-c_{\scaleto{\text{Ag}}{4.2pt}}) \right] + \widetilde{\Omega}_{a} k_{\text{B}}Tc_{\scaleto{\text{Ag}}{4.2pt}}(1-c_{\scaleto{\text{Ag}}{4.2pt}})
\end{eqnarray}
where $T =  310.15 K$ and $\widetilde{\Omega}_{a}=5.6$. The first-derivative of $g^{h}$ yields two stable phase compositions at $c_{\scaleto{\text{Ag}}{4.5pt}, 1}=0.003845$ and $c_{\scaleto{\text{Ag}}{4.5pt}, 2}=0.996155$.
The diffusional chemical potential $\mu^{h}_{\scaleto{\text{Ag}}{4.2pt}}$\ of Li$^{+}$ in inserted into SVO due to Ag$^{+}$ reduction is, 
\begin{eqnarray}\label{eq:105}
\mu^{h}_{\scaleto{\text{Ag}}{4.2pt}}(c_{\scaleto{\text{Ag}}{4.2pt}}) = \frac{\partial g}{\partial c_{\scaleto{\text{Ag}}{4.2pt}}} 
= k_{\text{B}}T \text{ln}(\frac{c_{\scaleto{\text{Ag}}{4.2pt}}}{1-c_{\scaleto{\text{Ag}}{4.2pt}}}) + \widetilde{\Omega}_{a} k_{\text{B}}T (1-2c_{\scaleto{\text{Ag}}{4.2pt}})
\end{eqnarray}
The second-derivative $g^{h}$ or the first-derivative of $\mu^{h}_{\scaleto{\text{Ag}}{4.2pt}}$ yields two metastable phase compositions at $c_{\scaleto{\text{Ag}}{4.5pt}, s_{1}}=0.09911$ and $c_{\scaleto{\text{Ag}}{4.5pt}, s_{2}}=0.90089$. 
The homogeneous OCV $\Delta\phi^{h}_{\text{Ag}}$ used in the simulation is thus,
\begin{eqnarray}\label{eq:106}
\Delta\phi^{h}_{\text{Ag}}(c_{\scaleto{\text{Ag}}{4.2pt}}) = 3.24 - \frac{\mu^{h}_{\scaleto{\text{Ag}}{4.2pt}}(c_{\scaleto{\text{Ag}}{4.2pt}})}{e} 
\end{eqnarray}
The thermodynamics of V$^{5+}$ reduction is described by using a separate OCV function $\Delta\phi^{eq}_{\text{V}}$, fitted directly as a function of its depth of discharge $c_{\scaleto{\text{V}}{4.2pt}}$ \cite{gomadam2007modeling},
\begin{eqnarray}\label{eq:107}
\Delta\phi^{eq}_{\text{V}}(c_{\scaleto{\text{V}}{4.2pt}}) = 0.823 + \text{exp}\left(-80c_{\scaleto{\text{V}}{4.2pt}}\right) + 
\frac{3.177 
+92.839c_{\scaleto{\text{V}}{4.2pt}}^2
+49.148c_{\scaleto{\text{V}}{4.2pt}}^4
-658.841c_{\scaleto{\text{V}}{4.2pt}}^6
+589.917c_{\scaleto{\text{V}}{4.2pt}}^8
}
{1 
+39.404c_{\scaleto{\text{V}}{4.2pt}}^2
-6.299c_{\scaleto{\text{V}}{4.2pt}}^4
-171.554c_{\scaleto{\text{V}}{4.2pt}}^6
+106.016c_{\scaleto{\text{V}}{4.2pt}}^8
+65.794c_{\scaleto{\text{V}}{4.2pt}}^{10}}
\end{eqnarray}
The reaction kinetics of Ag$^{+}$ and V$^{5+}$ reduction used are,
\begin{eqnarray}\label{eq:108}
i_{\text{\scaleto{\text{Ag}}{4.2pt}}} = j_{\scaleto{\text{Ag}}{4.2pt}} e = k_{\scaleto{\text{Ag}}{4.2pt}} (c^{l})^{0.5} c_{\text{\scaleto{\text{Ag}}{4.2pt}}}^{0.1}(1-c_{\text{\scaleto{\text{Ag}}{4.2pt}}})^{5.5}
\left[\text{exp}\left(-\frac{\alpha_{\text{\scaleto{\text{Ag}}{4.2pt}}} e \eta^{\text{eff}}_{\text{\scaleto{\text{Ag}}{4.2pt}}}}{k_{\text{B}}T}\right) - \text{exp}\left(\frac{(1-\alpha_{\text{\scaleto{\text{Ag}}{4.2pt}}}) e \eta^{\text{eff}}_{\text{\scaleto{\text{Ag}}{4.2pt}}}}{k_{\text{B}}T}\right) \right]
\end{eqnarray}
\begin{eqnarray}\label{eq:109}
i_{\text{\scaleto{\text{V}}{4.2pt}}} = j_{\scaleto{\text{V}}{4.2pt}} e = k_{\scaleto{\text{V}}{4.2pt}} (c^{l})^{0.5} c_{\text{\scaleto{\text{V}}{4.2pt}}}^{0.5}(1-c_{\text{\scaleto{\text{V}}{4.2pt}}})^{0.5}
\left[\text{exp}\left(-\frac{\alpha_{\text{\scaleto{\text{V}}{4.2pt}}} e \eta^{\text{eff}}_{\text{\scaleto{\text{V}}{4.2pt}}}}{k_{\text{B}}T}\right) - \text{exp}\left(\frac{(1-\alpha_{\text{\scaleto{\text{V}}{4.2pt}}}) e \eta^{\text{eff}}_{\text{\scaleto{\text{V}}{4.2pt}}}}{k_{\text{B}}T}\right) \right]
\end{eqnarray}
The separator used has thickness 50$\upmu$m, porosity 0.4, and tortuosity $\epsilon^{-0.6}$, and is discretized into 10 finite volumes. The other key parameters used in Hybrid-MPET model of Li/SVO can be seen in Table \ref{table:SVOelectrode} and many reference Gomadam et al. \cite{gomadam2007modeling},
\renewcommand{\thetable}{2}
\renewcommand{\arraystretch}{1.2}
\newcolumntype{P}[1]{>{\centering\arraybackslash}p{#1}}
\begin{table}[!h]
  \caption{Geometric and materials properties used in simulating SVO hybrid porous electrode}
  \centering
  \begin{tabular}{ c c c c c}
  \hline
   Parameter &Description & Unit & \multicolumn{2}{c}{SVO Hybrid Porous Electrode}
   \\
  \hline
    $L$ & thickness & mm &    \hspace{+55pt} 2.6 
    \\
    $\epsilon$ & porosity & - & \hspace{+55pt} 0.20    
    \\
    $P_{L}$ & active solid volume fraction & - & \hspace{+55pt} 0.95 
    \\
    $\tau$ & tortuosity & - & \hspace{+55pt} $\epsilon^{-0.6}$
    \\

  \hline
        Parameter &Description & Unit & Ag & V \\
  \hline
    $R$ & particle radius & $ \upmu$m  & \multicolumn{2}{c}{ \hspace{+5pt} $\mathcal{N}$(1 , 0.3)}
    \\
    $h$ & particle length & $ \upmu$m  & \multicolumn{2}{c}{ \hspace{+5pt} 20}
    \\
    $\rho$ & volumetric site density & mol m$^{-3}$ & 16107  & 32215   \\
    $\alpha$ & symmetry coefficient & - & 0.5 & 0.5
    \\
    $k$ & rate constant & A m$^{-2}$ & $2\times10^{-4}$ & $7\times10^{-1}$
    \\
    \hline
    \end{tabular}
  \label{table:SVOelectrode}
\end{table}

Adopting Eq. \ref{eq:4}, \ref{eq:6}, and \ref{eq:9} for SVO, and considering $\rho_{\scaleto{\text{Ag}}{4.2pt}} = \frac{1}{3}\rho_{\scaleto{\text{SVO}}{3.2pt}}$, $\rho_{\scaleto{\text{V}}{3.2pt}} = \frac{2}{3}\rho_{\scaleto{\text{SVO}}{3.2pt}}$, we have the following intra-particle scale equations,
\begin{eqnarray}\label{eq:22}
\overline{c}_{n,p} = \frac{1}{3}\overline{c}_{n, p, \scaleto{\text{Ag}}{4.5pt}}  + \frac{2}{3}\overline{c}_{n, p, \scaleto{\text{V}}{3.5pt}}
\end{eqnarray}
\begin{eqnarray}\label{eq:23}
R^{V}_{n} = -(1-\epsilon)P_{L}\sum_{p = 1}^{P} \widetilde{V}_{n, p}   \left( \frac{1}{3} \frac{ \partial \overline{c}_{n, p, \scaleto{\text{Ag}}{4.5pt}}}{\partial t} + \frac{2}{3} \frac{ \partial \overline{c}_{n, p, \scaleto{\text{V}}{3.5pt}}}{\partial t}\right)
\end{eqnarray}
where again $\widetilde{V}_{n, p} = \frac{V_{n, p}}{\sum_{p = 1}^{P}V_{n, p}} $ is volume fraction of particle $p$ in finite volume $n$.

\section*{Appendix C}
$\Delta\phi^{eq}_{\text{CF}}$ is extracted from experimental data by assuming Tafel kinetics, and fitted directly as a function of its depth of discharge $c_{\scaleto{\text{CF}}{3.5pt}}$,
\begin{multline}\label{eq:110}
\Delta\phi^{eq}_{\text{CF}}(c_{\scaleto{\text{CF}}{3.5pt}}) = -0.077
-0.047\text{ln}\left(c_{\scaleto{\text{CF}}{3.5pt}}(1-c_{\scaleto{\text{CF}}{3.5pt}})^{\frac{1}{3}}\right) \\
+\frac{
5.566
-1476.243c_{\scaleto{\text{CF}}{3.5pt}}
240158.880c_{\scaleto{\text{CF}}{3.5pt}}^2
-116028.195c_{\scaleto{\text{CF}}{3.5pt}}^3
-166268.131c_{\scaleto{\text{CF}}{3.5pt}}^4
+4436.904c_{\scaleto{\text{CF}}{3.5pt}}^5
}{
1
-404.085c_{\scaleto{\text{CF}}{3.5pt}}
+ 75655.382c_{\scaleto{\text{CF}}{3.5pt}}^2
-32080.426c_{\scaleto{\text{CF}}{3.5pt}}^3
-55461.302c_{\scaleto{\text{CF}}{3.5pt}}^4
} 
\end{multline}
The reaction kinetics of carbon monofluoride is,
\begin{eqnarray}\label{eq:108}
i_{\text{\scaleto{\text{CF}}{3.5pt}}} = j_{\scaleto{\text{CF}}{3.5pt}} e = k_{\scaleto{\text{CF}}{3.5pt}}
c_{\text{\scaleto{\text{CF}}{3.5pt}}}(1-c_{\text{\scaleto{\text{CF}}{3.5pt}}})
\left[\text{exp}\left(-\frac{\alpha_{\text{\scaleto{\text{CF}}{3.5pt}}} e \eta^{\text{eff}}_{\text{\scaleto{\text{CF}}{3.5pt}}}}{k_{\text{B}}T}\right)  \right]
\end{eqnarray}
The separator used has thickness 50$\upmu$m, porosity 0.4, and tortuosity $\epsilon^{-1}$, and is discretized into 10 finite volumes. The other key parameters used in Hybrid-MPET model of Li/CF$_{\text{x}}$-SVO can be seen in Table \ref{table:CFXSVOelectrode} and many reference Gomadam et al. \cite{gomadam2007modeling},
\renewcommand{\thetable}{3}
\renewcommand{\arraystretch}{1.2}
\newcolumntype{P}[1]{>{\centering\arraybackslash}p{#1}}
\begin{table}[!h]
  \caption{Geometric and materials properties used in simulating CF$_{\text{x}}$-SVO hybrid porous electrode}
  \centering
  \begin{tabular}{ccccccc}
  \hline
   Parameter &Description & Unit & HB1 & HB2 & HB3 & HB4 
   \\
  \hline
    $L$ & thickness & mm & 2.1 & 2.1 & 2.1 & 2.6 
    \\
    $\epsilon$ & porosity & - & 0.33 & 0.33 & 0.33 & 0.33
    \\
    $P_{L}$ & active solid volume fraction & - & 0.88 & 0.85 & 0.84 & 0.85 
    \\
    $\tau$ & tortuosity & - & $\epsilon^{-1}$ & $\epsilon^{-1}$ & $\epsilon^{-1}$ & $\epsilon^{-1}$
    \\
    $\widetilde{Q}_{SVO} $ & capacity fraction of silicon & -  & 0.128 & 0.128 & 0.333  & 0.128 
    \\
    $\widetilde{Q}_{CF}$ & capacity fraction of graphite & -  & 0.872 & 0.872 & 0.667 & 0.872 
    \\
      
  \hline
        Parameter &Description & Unit & \multicolumn{2}{c}{\hspace{+0pt} Ag} & V & CF$_{\text{x}}$ \\
  \hline
    $R$ & particle radius & $ \upmu$m  & \multicolumn{3}{c}{\hspace{+10pt} $\mathcal{N}$(1, 0.3)}  & $\mathcal{N}$(10, 0.5)
    \\
    $h$ & particle length & $ \upmu$m  & \multicolumn{3}{c}{\hspace{+10pt} 20}  & -
    \\
    $\rho$ & volumetric site density & mol m$^{-3}$ & \multicolumn{2}{c}{16107} & 32215  & 88739 \\
    $\alpha$ & symmetry coefficient & - & \multicolumn{2}{c}{0.5} & 0.5 & 0.57
    \\
    $k$ & rate constant & A m$^{-2}$ & \multicolumn{2}{c}{3.5$\times10^{-5}$} & 1$\times10^{-3}$ & 5$\times10^{-4}$
    \\
    \hline
    \end{tabular}
  \label{table:CFXSVOelectrode}
\end{table}

Since CF$_{\text{x}}$-SVO is a hybrid electrode at both intra-particle and inter-particle scale, we adopt Eq. \ref{eq:4}, \ref{eq:6} and \ref{eq:13}, \ref{eq:16}. Again, since $\rho_{\scaleto{\text{Ag}}{4.2pt}} = \frac{1}{3}\rho_{\scaleto{\text{SVO}}{3.2pt}}$, $\rho_{\scaleto{\text{V}}{3.2pt}} = \frac{2}{3}\rho_{\scaleto{\text{SVO}}{3.2pt}}$, we thus have,
\begin{eqnarray}\label{eq:24}
\overline{c}^{corr}_{n} =\frac{\sum_{p_{\scaleto{\text{SVO}}{2.7pt}} = 1}^{P_{\scaleto{\text{SVO}}{2.7pt}}}\rho_{\scaleto{\text{SVO}}{3.2pt}}V_{n, p_{\scaleto{\text{SVO}}{2.7pt}}}f_{n,\scaleto{\text{SVO}}{3.2pt}}(\frac{1}{3}\overline{c}_{n,p_{\scaleto{\text{SVO}}{2.7pt}}, \scaleto{\text{Ag}}{4.5pt}} + \frac{2}{3}\overline{c}_{n,p_{\scaleto{\text{SVO}}{2.7pt}}, \scaleto{\text{V}}{3.5pt}}) + \sum_{p_{\scaleto{\text{CF}}{2.7pt}} = 1}^{P_{\scaleto{\text{CF}}{2.7pt}}}\rho_{\scaleto{\text{CF}}{3.2pt}}V_{n, p_{\scaleto{\text{CF}}{2.7pt}}}f_{n,\scaleto{\text{CF}}{3.2pt}}\overline{c}_{n,p_{\scaleto{\text{CF}}{2.7pt}}}
}{\sum_{p_{\scaleto{\text{SVO}}{2.7pt}} = 1}^{P_{\scaleto{\text{SVO}}{2.7pt}}}\rho_{\scaleto{\text{SVO}}{3.2pt}}V_{n, p_{\scaleto{\text{SVO}}{2.7pt}}}f_{n,\scaleto{\text{SVO}}{3.2pt}} +\sum_{p_{\scaleto{\text{CF}}{2.7pt}} = 1}^{P_{\scaleto{\text{CF}}{2.7pt}}}\rho_{\scaleto{\text{CF}}{3.2pt}}V_{n, p_{\scaleto{\text{CF}}{2.7pt}}}f_{n,\scaleto{\text{CF}}{3.2pt}}}
\end{eqnarray}
\begin{eqnarray}\label{eq:25}
R^{V, corr}_{n} = -(1-\epsilon)P_{L}\left(
\sum_{p_{\scaleto{\text{SVO}}{3.2pt}} = 1}^{P_{\scaleto{\text{SVO}}{3.2pt}}} \widetilde{V}_{n, p_{\scaleto{\text{SVO}}{2.7pt}}} \left(\frac{1}{3}\frac{\partial \overline{c}_{n,p_{\scaleto{\text{SVO}}{2.7pt}}, \scaleto{\text{Ag}}{4.5pt}}}{\partial t} + \frac{2}{3} \frac{\partial \overline{c}_{n,p_{\scaleto{\text{SVO}}{2.7pt}}, \scaleto{\text{V}}{3.5pt}}}{\partial t} \right)
+
\sum_{p_{\scaleto{\text{CF}}{3.2pt}} = 1}^{P_{\scaleto{\text{CF}}{3.2pt}}} \widetilde{V}_{n, p_{\scaleto{\text{CF}}{2.7pt}}} \frac{ \partial \overline{c}_{n,p_{\scaleto{\text{CF}}{2.7pt}}}}{\partial t}
\right) 
\end{eqnarray} 
where the corrected volume fractions are
\begin{eqnarray}\label{eq:26}
\widetilde{V}_{n, p_{\scaleto{\text{SVO}}{2.7pt}}} = \frac{V_{n, p_{\scaleto{\text{SVO}}{2.7pt}}}f_{n,\scaleto{\text{SVO}}{3.2pt}}}{\sum_{p_{\scaleto{\text{SVO}}{2.7pt}} = 1}^{P_{\scaleto{\text{SVO}}{2.7pt}}}V_{n,p_{\scaleto{\text{SVO}}{2.7pt}}}f_{n,\scaleto{\text{SVO}}{3.2pt}} 
+
\sum_{p_{\scaleto{\text{CF}}{2.7pt}} = 1}^{P_{\scaleto{\text{CF}}{2.7pt}}}V_{n,p_{\scaleto{\text{CF}}{2.7pt}}}f_{n,\scaleto{\text{CF}}{3.2pt}} }
\end{eqnarray} 
\begin{eqnarray}\label{eq:27}
\widetilde{V}_{n, p_{\scaleto{\text{CF}}{2.7pt}}} = \frac{V_{n, p_{\scaleto{\text{CF}}{2.7pt}}}f_{n,\scaleto{\text{CF}}{3.2pt}}}{\sum_{p_{\scaleto{\text{SVO}}{2.7pt}} = 1}^{P_{\scaleto{\text{SVO}}{2.7pt}}}V_{n,p_{\scaleto{\text{SVO}}{2.7pt}}}f_{n,\scaleto{\text{SVO}}{3.2pt}} 
+
\sum_{p_{\scaleto{\text{CF}}{2.7pt}} = 1}^{P_{\scaleto{\text{CF}}{2.7pt}}}V_{n,p_{\scaleto{\text{CF}}{2.7pt}}}f_{n,\scaleto{\text{CF}}{3.2pt}} }
\end{eqnarray} 

\section*{Supplementary Material}
Supplementary Material is available with the electronic version of the article on the journal's website. 

\section*{Competing interests}
The authors declare no competing interests.

\vspace{+20pt}

\printbibliography

@misc{hybridMPETGithub,
  author = {Liang, Qiaohao},
  title = {Hybrid-MPET: an open-source simulation software for hybrid electrode batteries},
  year = {2023},
  publisher = {GitHub},
  howpublished = {at \url{https://github.com/HarryQL/Hybrid-MPET}},
}

@article{pinson2012theory,
  title={Theory of SEI formation in rechargeable batteries: capacity fade, accelerated aging and lifetime prediction},
  author={Pinson, Matthew B and Bazant, Martin Z},
  journal={Journal of the Electrochemical Society},
  volume={160},
  number={2},
  pages={A243},
  year={2012},
  publisher={IOP Publishing}
}

@article{thomas2017situ,
  title={In situ observation and mathematical modeling of lithium distribution within graphite},
  author={Thomas-Alyea, Karen E and Jung, Changhoon and Smith, Raymond B and Bazant, Martin Z},
  journal={Journal of The Electrochemical Society},
  volume={164},
  number={11},
  pages={E3063},
  year={2017},
  publisher={IOP Publishing}
}

@article{zhuang2022theory,
  title={Theory of layered-oxide cathode degradation in Li-ion batteries by oxidation-induced cation disorder},
  author={Zhuang, Debbie and Bazant, Martin Z},
  journal={Journal of The Electrochemical Society},
  volume={169},
  number={10},
  pages={100536},
  year={2022},
  publisher={IOP Publishing}
}

@article{finegan2020spatial,
  title={Spatial dynamics of lithiation and lithium plating during high-rate operation of graphite electrodes},
  author={Finegan, Donal P and Quinn, Alexander and Wragg, David S and Colclasure, Andrew M and Lu, Xuekun and Tan, Chun and Heenan, Thomas MM and Jervis, Rhodri and Brett, Dan JL and Das, Supratim and others},
  journal={Energy \& Environmental Science},
  volume={13},
  number={8},
  pages={2570--2584},
  year={2020},
  publisher={Royal Society of Chemistry}
}

@article{dose2018capacity,
  title={Capacity fade in high energy silicon-graphite electrodes for lithium-ion batteries},
  author={Dose, WM and Piernas-Mu{\~n}oz, MJ and Maroni, VA and Trask, SE and Bloom, I and Johnson, CS},
  journal={Chemical Communications},
  volume={54},
  number={29},
  pages={3586--3589},
  year={2018},
  publisher={Royal Society of Chemistry}
}

@article{dhillon2021modelling,
  title={Modelling capacity fade in silicon-graphite composite electrodes for lithium-ion batteries},
  author={Dhillon, Shweta and Hern{\'a}ndez, Guiomar and Wagner, Nils P and Svensson, Ann Mari and Brandell, Daniel},
  journal={Electrochimica Acta},
  volume={377},
  pages={138067},
  year={2021},
  publisher={Elsevier}
}

@article{jana2022physics,
  title={Physics-based, reduced order degradation model of lithium-ion batteries},
  author={Jana, Aniruddha and Mitra, A Surya and Das, Supratim and Chueh, William C and Bazant, Martin Z and Garc{\'\i}a, R Edwin},
  journal={Journal of Power Sources},
  volume={545},
  pages={231900},
  year={2022},
  publisher={Elsevier}
}

@article{arico2005nanostructured,
  title={Nanostructured materials for advanced energy conversion and storage devices},
  author={Arico, Antonino Salvatore and Bruce, Peter and Scrosati, Bruno and Tarascon, Jean-Marie and Van Schalkwijk, Walter},
  journal={Nature materials},
  volume={4},
  number={5},
  pages={366--377},
  year={2005},
  publisher={Nature Publishing Group UK London}
}

@article{chu2017path,
  title={The path towards sustainable energy},
  author={Chu, Steven and Cui, Yi and Liu, Nian},
  journal={Nature materials},
  volume={16},
  number={1},
  pages={16--22},
  year={2017},
  publisher={Nature Publishing Group UK London}
}

@article{tarascon2001issues,
  title={Issues and challenges facing rechargeable lithium batteries},
  author={Tarascon, J-M and Armand, Michel},
  journal={nature},
  volume={414},
  number={6861},
  pages={359--367},
  year={2001},
  publisher={Nature Publishing Group UK London}
}

@article{padhi1997phospho,
  title={Phospho-olivines as positive-electrode materials for rechargeable lithium batteries},
  author={Padhi, Akshaya K and Nanjundaswamy, Kirakodu S and Goodenough, John B},
  journal={Journal of the electrochemical society},
  volume={144},
  number={4},
  pages={1188},
  year={1997},
  publisher={IOP Publishing}
}

@article{guo2016li,
  title={Li intercalation into graphite: direct optical imaging and Cahn--Hilliard reaction dynamics},
  author={Guo, Yinsheng and Smith, Raymond B and Yu, Zhonghua and Efetov, Dmitri K and Wang, Junpu and Kim, Philip and Bazant, Martin Z and Brus, Louis E},
  journal={The journal of physical chemistry letters},
  volume={7},
  number={11},
  pages={2151--2156},
  year={2016},
  publisher={ACS Publications}
}

@inproceedings{pillot2017rechargeable,
  title={The rechargeable battery market and main trends 2016--2025},
  author={Pillot, Christophe},
  booktitle={Proceedings of the 33rd Annual International Battery Seminar \& Exhibit, Fort Lauderdale, FL, USA},
  volume={20},
  year={2017}
}

@article{whittingham2004lithium,
  title={Lithium batteries and cathode materials},
  author={Whittingham, M Stanley},
  journal={Chemical reviews},
  volume={104},
  number={10},
  pages={4271--4302},
  year={2004},
  publisher={ACS Publications}
}

@article{chakraborty2020layered,
  title={Layered cathode materials for lithium-ion batteries: review of computational studies on LiNi1--x--y Co x Mn y O2 and LiNi1--x--y Co x Al y O2},
  author={Chakraborty, Arup and Kunnikuruvan, Sooraj and Kumar, Sandeep and Markovsky, Boris and Aurbach, Doron and Dixit, Mudit and Major, Dan Thomas},
  journal={Chemistry of Materials},
  volume={32},
  number={3},
  pages={915--952},
  year={2020},
  publisher={ACS Publications}
}

@article{thackeray2021layered,
  title={Layered Li--Ni--Mn--Co oxide cathodes},
  author={Thackeray, Michael M and Amine, Khalil},
  journal={Nature Energy},
  volume={6},
  number={9},
  pages={933--933},
  year={2021},
  publisher={Nature Publishing Group UK London}
}

@article{park2021fictitious,
  title={Fictitious phase separation in Li layered oxides driven by electro-autocatalysis},
  author={Park, Jungjin and Zhao, Hongbo and Kang, Stephen Dongmin and Lim, Kipil and Chen, Chia-Chin and Yu, Young-Sang and Braatz, Richard D and Shapiro, David A and Hong, Jihyun and Toney, Michael F and others},
  journal={Nature Materials},
  volume={20},
  number={7},
  pages={991--999},
  year={2021},
  publisher={Nature Publishing Group UK London}
}

@article{xia2018designing,
  title={Designing principle for Ni-rich cathode materials with high energy density for practical applications},
  author={Xia, Yu and Zheng, Jianming and Wang, Chongmin and Gu, Meng},
  journal={Nano Energy},
  volume={49},
  pages={434--452},
  year={2018},
  publisher={Elsevier}
}

@article{radin2017narrowing,
  title={Narrowing the gap between theoretical and practical capacities in Li-ion layered oxide cathode materials},
  author={Radin, Maxwell D and Hy, Sunny and Sina, Mahsa and Fang, Chengcheng and Liu, Haodong and Vinckeviciute, Julija and Zhang, Minghao and Whittingham, M Stanley and Meng, Y Shirley and Van der Ven, Anton},
  journal={Advanced Energy Materials},
  volume={7},
  number={20},
  pages={1602888},
  year={2017},
  publisher={Wiley Online Library}
}

@article{kang2009battery,
  title={Battery materials for ultrafast charging and discharging},
  author={Kang, Byoungwoo and Ceder, Gerbrand},
  journal={Nature},
  volume={458},
  number={7235},
  pages={190--193},
  year={2009},
  publisher={Nature Publishing Group UK London}
}

@article{gao2021interplay,
  title={Interplay of lithium intercalation and plating on a single graphite particle},
  author={Gao, Tao and Han, Yu and Fraggedakis, Dimitrios and Das, Supratim and Zhou, Tingtao and Yeh, Che-Ning and Xu, Shengming and Chueh, William C and Li, Ju and Bazant, Martin Z},
  journal={Joule},
  volume={5},
  number={2},
  pages={393--414},
  year={2021},
  publisher={Elsevier}
}

@article{yu2013hybrid,
  title={Hybrid nanostructured materials for high-performance electrochemical capacitors},
  author={Yu, Guihua and Xie, Xing and Pan, Lijia and Bao, Zhenan and Cui, Yi},
  journal={Nano Energy},
  volume={2},
  number={2},
  pages={213--234},
  year={2013},
  publisher={Elsevier}
}

@article{simon2008materials,
  title={Materials for electrochemical capacitors},
  author={Simon, Patrice and Gogotsi, Yury},
  journal={Nature materials},
  volume={7},
  number={11},
  pages={845--854},
  year={2008},
  publisher={Nature Publishing Group UK London}
}

@article{andre2015future,
  title={Future generations of cathode materials: an automotive industry perspective},
  author={Andre, Dave and Kim, Sung-Jin and Lamp, Peter and Lux, Simon Franz and Maglia, Filippo and Paschos, Odysseas and Stiaszny, Barbara},
  journal={Journal of Materials Chemistry A},
  volume={3},
  number={13},
  pages={6709--6732},
  year={2015},
  publisher={Royal Society of Chemistry}
}

@article{amine2005advanced,
  title={Advanced cathode materials for high-power applications},
  author={Amine, K and Liu, J and Belharouak, I and Kang, S-H and Bloom, I and Vissers, D and Henriksen, G},
  journal={Journal of power sources},
  volume={146},
  number={1-2},
  pages={111--115},
  year={2005},
  publisher={Elsevier}
}

@article{heo2022amorphous,
  title={Amorphous iron fluorosulfate as a high-capacity cathode utilizing combined intercalation and conversion reactions with unexpectedly high reversibility},
  author={Heo, Jaehoon and Jung, Sung-Kyun and Hwang, Insang and Cho, Sung-Pyo and Eum, Donggun and Park, Hyeokjun and Song, Jun-Hyuk and Yu, Seungju and Oh, Kyungbae and Kwon, Giyun and others},
  journal={Nature Energy},
  pages={1--10},
  year={2022},
  publisher={Nature Publishing Group UK London}
}

@article{cabana2010beyond,
  title={Beyond intercalation-based Li-ion batteries: the state of the art and challenges of electrode materials reacting through conversion reactions},
  author={Cabana, Jordi and Monconduit, Laure and Larcher, Dominique and Palacin, M Rosa},
  journal={Advanced materials},
  volume={22},
  number={35},
  pages={E170--E192},
  year={2010},
  publisher={Wiley Online Library}
}

@article{wu2017conversion,
  title={Conversion cathodes for rechargeable lithium and lithium-ion batteries},
  author={Wu, Feixiang and Yushin, Gleb},
  journal={Energy \& Environmental Science},
  volume={10},
  number={2},
  pages={435--459},
  year={2017},
  publisher={Royal Society of Chemistry}
}

@article{chikkannanavar2014review,
  title={A review of blended cathode materials for use in Li-ion batteries},
  author={Chikkannanavar, Satishkumar B and Bernardi, Dawn M and Liu, Lingyun},
  journal={Journal of Power Sources},
  volume={248},
  pages={91--100},
  year={2014},
  publisher={Elsevier}
}

@article{gao2009eliminating,
  title={Eliminating the irreversible capacity loss of high capacity layered Li [Li0. 2Mn0. 54Ni0. 13Co0. 13] O2 cathode by blending with other lithium insertion hosts},
  author={Gao, J and Manthiram, Arumugam},
  journal={Journal of Power Sources},
  volume={191},
  number={2},
  pages={644--647},
  year={2009},
  publisher={Elsevier}
}

@article{kitao2004high,
  title={High-temperature storage performance of Li-ion batteries using a mixture of Li-Mn spinel and Li-Ni-Co-Mn oxide as a positive electrode material},
  author={Kitao, Hideki and Fujihara, Toyoki and Takeda, Kazuhisa and Nakanishi, Naoya and Nohma, Toshiyuki},
  journal={Electrochemical and solid-state letters},
  volume={8},
  number={2},
  pages={A87},
  year={2004},
  publisher={IOP Publishing}
}

@article{albertus2009experiments,
  title={Experiments on and modeling of positive electrodes with multiple active materials for lithium-ion batteries},
  author={Albertus, Paul and Christensen, Jake and Newman, John},
  journal={Journal of The Electrochemical Society},
  volume={156},
  number={7},
  pages={A606},
  year={2009},
  publisher={IOP Publishing}
}

@article{lee2011alf3,
  title={AlF3-coated LiCoO2 and Li [Ni1/3Co1/3Mn1/3] O2 blend composite cathode for lithium ion batteries},
  author={Lee, Ki-Soo and Myung, Seung-Taek and Kim, Dong-Won and Sun, Yang-Kook},
  journal={Journal of Power Sources},
  volume={196},
  number={16},
  pages={6974--6977},
  year={2011},
  publisher={Elsevier}
}

@article{gallagher2011xli2mno3,
  title={xLi2MnO3{\textperiodcentered}(1- x) LiMO2 blended with LiFePO4 to achieve high energy density and pulse power capability},
  author={Gallagher, Kevin G and Kang, Sun-Ho and Park, Sei Ung and Han, Soo Young},
  journal={Journal of Power Sources},
  volume={196},
  number={22},
  pages={9702--9707},
  year={2011},
  publisher={Elsevier}
}

@article{kang2009enhancing,
  title={Enhancing the rate capability of high capacity xLi2MnO3{\textperiodcentered}(1- x) LiMO2 (M= Mn, Ni, Co) electrodes by Li--Ni--PO4 treatment},
  author={Kang, Sun-Ho and Thackeray, Michael M},
  journal={Electrochemistry Communications},
  volume={11},
  number={4},
  pages={748--751},
  year={2009},
  publisher={Elsevier}
}

@article{thackeray2007li,
  title={Li 2 MnO 3-stabilized LiMO 2 (M= Mn, Ni, Co) electrodes for lithium-ion batteries},
  author={Thackeray, Michael M and Kang, Sun-Ho and Johnson, Christopher S and Vaughey, John T and Benedek, Roy and Hackney, SA},
  journal={Journal of Materials chemistry},
  volume={17},
  number={30},
  pages={3112--3125},
  year={2007},
  publisher={Royal Society of Chemistry}
}

@article{johnson2008synthesis,
  title={Synthesis, characterization and electrochemistry of lithium battery electrodes: x Li2MnO3{\textperiodcentered}(1- x) LiMn0. 333Ni0. 333Co0. 333O2 (0$\le$ x$\le$ 0.7)},
  author={Johnson, Christopher S and Li, Naichao and Lefief, Christina and Vaughey, John T and Thackeray, Michael M},
  journal={Chemistry of Materials},
  volume={20},
  number={19},
  pages={6095--6106},
  year={2008},
  publisher={ACS Publications}
}

@article{schmidt2001future,
  title={The future of lithium and lithium-ion batteries in implantable medical devices},
  author={Schmidt, Craig L and Skarstad, Paul M},
  journal={Journal of power sources},
  volume={97},
  pages={742--746},
  year={2001},
  publisher={Elsevier}
}

@article{gan2005dual,
  title={Dual-chemistry cathode system for high-rate pulse applications},
  author={Gan, Hong and Rubino, Robert S and Takeuchi, Esther S},
  journal={Journal of power sources},
  volume={146},
  number={1-2},
  pages={101--106},
  year={2005},
  publisher={Elsevier}
}

@article{chen2006hybrid,
  title={Hybrid cathode lithium batteries for implantable medical applications},
  author={Chen, Kaimin and Merritt, Donald R and Howard, William G and Schmidt, Craig L and Skarstad, Paul M},
  journal={Journal of power sources},
  volume={162},
  number={2},
  pages={837--840},
  year={2006},
  publisher={Elsevier}
}

@article{ai2022composite,
  title={A composite electrode model for lithium-ion batteries with silicon/graphite negative electrodes},
  author={Ai, Weilong and Kirkaldy, Niall and Jiang, Yang and Offer, Gregory and Wang, Huizhi and Wu, Billy},
  journal={Journal of Power Sources},
  volume={527},
  pages={231142},
  year={2022},
  publisher={Elsevier}
}

@article{schmitt2021change,
  title={Change in the half-cell open-circuit potential curves of silicon--graphite and nickel-rich lithium nickel manganese cobalt oxide during cycle aging},
  author={Schmitt, Julius and Schindler, Markus and Jossen, Andreas},
  journal={Journal of Power Sources},
  volume={506},
  pages={230240},
  year={2021},
  publisher={Elsevier}
}

@article{li2021diverting,
  title={Diverting exploration of silicon anode into practical way: a review focused on silicon-graphite composite for lithium ion batteries},
  author={Li, Peng and Kim, Hun and Myung, Seung-Taek and Sun, Yang-Kook},
  journal={Energy Storage Materials},
  volume={35},
  pages={550--576},
  year={2021},
  publisher={Elsevier}
}

@article{yoshio2006silicon,
  title={Silicon/graphite composites as an anode material for lithium ion batteries},
  author={Yoshio, Masaki and Tsumura, Takaaki and Dimov, Nikolay},
  journal={Journal of power sources},
  volume={163},
  number={1},
  pages={215--218},
  year={2006},
  publisher={Elsevier}
}

@article{jin2017challenges,
  title={Challenges and recent progress in the development of Si anodes for lithium-ion battery},
  author={Jin, Yan and Zhu, Bin and Lu, Zhenda and Liu, Nian and Zhu, Jia},
  journal={Advanced Energy Materials},
  volume={7},
  number={23},
  pages={1700715},
  year={2017},
  publisher={Wiley Online Library}
}

@article{chen2020development,
  title={Development of experimental techniques for parameterization of multi-scale lithium-ion battery models},
  author={Chen, Chang-Hui and Planella, Ferran Brosa and O’regan, Kieran and Gastol, Dominika and Widanage, W Dhammika and Kendrick, Emma},
  journal={Journal of The Electrochemical Society},
  volume={167},
  number={8},
  pages={080534},
  year={2020},
  publisher={IOP Publishing}
}

@article{li2007situ,
  title={An in situ X-ray diffraction study of the reaction of Li with crystalline Si},
  author={Li, Jing and Dahn, JR},
  journal={Journal of The Electrochemical Society},
  volume={154},
  number={3},
  pages={A156},
  year={2007},
  publisher={IOP Publishing}
}

@article{liu2011situ,
  title={In situ TEM electrochemistry of anode materials in lithium ion batteries},
  author={Liu, Xiao Hua and Huang, Jian Yu},
  journal={Energy \& Environmental Science},
  volume={4},
  number={10},
  pages={3844--3860},
  year={2011},
  publisher={Royal Society of Chemistry}
}

@article{smith2017multiphase,
  title={Multiphase porous electrode theory},
  author={Smith, Raymond B and Bazant, Martin Z},
  journal={Journal of The Electrochemical Society},
  volume={164},
  number={11},
  pages={E3291},
  year={2017},
  publisher={IOP Publishing}
}

@article{harris2013effects,
  title={Effects of Inhomogeneities-Nanoscale to Mesoscale-on the Durability of Li-Ion Batteries},
  author={Harris, Stephen J and Lu, Peng},
  journal={The Journal of Physical Chemistry C},
  volume={117},
  number={13},
  pages={6481--6492},
  year={2013},
  publisher={ACS Publications}
}

@article{li2018fluid,
  title={Fluid-enhanced surface diffusion controls intraparticle phase transformations},
  author={Li, Yiyang and Chen, Hungru and Lim, Kipil and Deng, Haitao D and Lim, Jongwoo and Fraggedakis, Dimitrios and Attia, Peter M and Lee, Sang Chul and Jin, Norman and Mo{\v{s}}kon, Jo{\v{z}}e and others},
  journal={Nature materials},
  volume={17},
  number={10},
  pages={915--922},
  year={2018},
  publisher={Nature Publishing Group UK London}
}

@article{li2014current,
  title={Current-induced transition from particle-by-particle to concurrent intercalation in phase-separating battery electrodes},
  author={Li, Yiyang and El Gabaly, Farid and Ferguson, Todd R and Smith, Raymond B and Bartelt, Norman C and Sugar, Joshua D and Fenton, Kyle R and Cogswell, Daniel A and Kilcoyne, AL David and Tyliszczak, Tolek and others},
  journal={Nature materials},
  volume={13},
  number={12},
  pages={1149--1156},
  year={2014},
  publisher={Nature Publishing Group UK London}
}

@article{chueh2013intercalation,
  title={Intercalation pathway in many-particle LiFePO4 electrode revealed by nanoscale state-of-charge mapping},
  author={Chueh, William C and El Gabaly, Farid and Sugar, Joshua D and Bartelt, Norman C and McDaniel, Anthony H and Fenton, Kyle R and Zavadil, Kevin R and Tyliszczak, Tolek and Lai, Wei and McCarty, Kevin F},
  journal={Nano letters},
  volume={13},
  number={3},
  pages={866--872},
  year={2013},
  publisher={ACS Publications}
}

@article{bazant2022learning,
  title={Learning heterogeneous reaction kinetics from X-ray movies pixel-by-pixel},
  author={Bazant, Martin and Zhao, Hongbo and Deng, Haitao and Cohen, Alexander and Lim, Jongwoo and Li, Yiyang and Fraggedakis, Dimitrios and Jiang, Benben and Storey, Brian and Chueh, William and others},
  year={2022}
}

@article{agrawal2022dynamic,
  title={Dynamic interplay between phase transformation instabilities and reaction heterogeneities in particulate intercalation electrodes},
  author={Agrawal, Shubham and Bai, Peng},
  journal={Cell Reports Physical Science},
  volume={3},
  number={5},
  pages={100854},
  year={2022},
  publisher={Elsevier}
}

@article{agrawal2021operando,
  title={Operando electrochemical kinetics in particulate porous electrodes by quantifying the mesoscale spatiotemporal heterogeneities},
  author={Agrawal, Shubham and Bai, Peng},
  journal={Advanced Energy Materials},
  volume={11},
  number={12},
  pages={2003344},
  year={2021},
  publisher={Wiley Online Library}
}

@article{ferguson2012nonequilibrium,
  title={Nonequilibrium thermodynamics of porous electrodes},
  author={Ferguson, Todd R and Bazant, Martin Z},
  journal={Journal of The Electrochemical Society},
  volume={159},
  number={12},
  pages={A1967},
  year={2012},
  publisher={IOP Publishing}
}

@article{bai2011suppression,
  title={Suppression of phase separation in LiFePO4 nanoparticles during battery discharge},
  author={Bai, Peng and Cogswell, Daniel A and Bazant, Martin Z},
  journal={Nano letters},
  volume={11},
  number={11},
  pages={4890--4896},
  year={2011},
  publisher={ACS Publications}
}

@article{ferguson2014phase,
  title={Phase transformation dynamics in porous battery electrodes},
  author={Ferguson, Todd R and Bazant, Martin Z},
  journal={Electrochimica Acta},
  volume={146},
  pages={89--97},
  year={2014},
  publisher={Elsevier}
}

@article{newman1975porous,
  title={Porous-electrode theory with battery applications},
  author={Newman, John and Tiedemann, William},
  journal={AIChE Journal},
  volume={21},
  number={1},
  pages={25--41},
  year={1975},
  publisher={Wiley Online Library}
}

@article{newman1962theoretical,
  title={Theoretical analysis of current distribution in porous electrodes},
  author={Newman, John S and Tobias, Charles W},
  journal={Journal of The Electrochemical Society},
  volume={109},
  number={12},
  pages={1183},
  year={1962},
  publisher={IOP Publishing}
}

@article{doyle1993modeling,
  title={Modeling of galvanostatic charge and discharge of the lithium/polymer/insertion cell},
  author={Doyle, Marc and Fuller, Thomas F and Newman, John},
  journal={Journal of the Electrochemical society},
  volume={140},
  number={6},
  pages={1526},
  year={1993},
  publisher={IOP Publishing}
}

@article{fuller1994simulation,
  title={Simulation and optimization of the dual lithium ion insertion cell},
  author={Fuller, Thomas F and Doyle, Marc and Newman, John},
  journal={Journal of the electrochemical society},
  volume={141},
  number={1},
  pages={1},
  year={1994},
  publisher={IOP Publishing}
}

@article{doyle1996comparison,
  title={Comparison of modeling predictions with experimental data from plastic lithium ion cells},
  author={Doyle, Marc and Newman, John and Gozdz, Antoni S and Schmutz, Caroline N and Tarascon, Jean-Marie},
  journal={Journal of the Electrochemical Society},
  volume={143},
  number={6},
  pages={1890},
  year={1996},
  publisher={IOP Publishing}
}

@article{torchio2016lionsimba,
  title={Lionsimba: a matlab framework based on a finite volume model suitable for li-ion battery design, simulation, and control},
  author={Torchio, Marcello and Magni, Lalo and Gopaluni, R Bhushan and Braatz, Richard D and Raimondo, Davide M},
  journal={Journal of The Electrochemical Society},
  volume={163},
  number={7},
  pages={A1192},
  year={2016},
  publisher={IOP Publishing}
}

@article{sulzer2021python,
  title={Python battery mathematical modelling (PyBaMM)},
  author={Sulzer, Valentin and Marquis, Scott G and Timms, Robert and Robinson, Martin and Chapman, S Jon},
  journal={Journal of Open Research Software},
  volume={9},
  number={1},
  year={2021},
  publisher={Ubiquity Press}
}

@article{berliner2021methods,
  title={Methods—PETLION: Open-source software for millisecond-scale porous electrode theory-based lithium-ion battery simulations},
  author={Berliner, Marc D and Cogswell, Daniel A and Bazant, Martin Z and Braatz, Richard D},
  journal={Journal of The Electrochemical Society},
  volume={168},
  number={9},
  pages={090504},
  year={2021},
  publisher={IOP Publishing}
}

@article{albertus2007introduction,
  title={Introduction to dualfoil 5.0},
  author={Albertus, Paul and Newman, Johon},
  journal={University of California Berkeley, Berkeley, CA, Tech. Rep},
  year={2007}
}

@article{bazant2013theory,
  title={Theory of chemical kinetics and charge transfer based on nonequilibrium thermodynamics},
  author={Bazant, Martin Z},
  journal={Accounts of chemical research},
  volume={46},
  number={5},
  pages={1144--1160},
  year={2013},
  publisher={ACS Publications}
}

@article{bazant2017thermodynamic,
  title={Thermodynamic stability of driven open systems and control of phase separation by electro-autocatalysis},
  author={Bazant, Martin Z},
  journal={Faraday discussions},
  volume={199},
  pages={423--463},
  year={2017},
  publisher={Royal Society of Chemistry}
}

@article{bazant2012phase,
  title={Phase-field theory of ion intercalation kinetics},
  author={Bazant, Martin Z},
  journal={arXiv preprint arXiv:1208.1587},
  year={2012}
}

@article{singh2008intercalation,
  title={Intercalation dynamics in rechargeable battery materials: General theory and phase-transformation waves in LiFePO4},
  author={Singh, Gogi K and Ceder, Gerbrand and Bazant, Martin Z},
  journal={Electrochimica Acta},
  volume={53},
  number={26},
  pages={7599--7613},
  year={2008},
  publisher={Elsevier}
}

@inproceedings{burch2008phase,
  title={Phase-transformation wave dynamics in LiFePO4},
  author={Burch, Damian and Singh, Gogi and Ceder, Gerbrand and Bazant, Martin Z},
  booktitle={Solid State Phenomena},
  volume={139},
  pages={95--100},
  year={2008},
  organization={Trans Tech Publ}
}

@article{bazant2009towards,
  title={Towards an understanding of induced-charge electrokinetics at large applied voltages in concentrated solutions},
  author={Bazant, Martin Z and Kilic, Mustafa Sabri and Storey, Brian D and Ajdari, Armand},
  journal={Advances in colloid and interface science},
  volume={152},
  number={1-2},
  pages={48--88},
  year={2009},
  publisher={Elsevier}
}

@article{burch2009size,
  title={Size-dependent spinodal and miscibility gaps for intercalation in nanoparticles},
  author={Burch, Damian and Bazant, Martin Z},
  journal={Nano letters},
  volume={9},
  number={11},
  pages={3795--3800},
  year={2009},
  publisher={ACS Publications}
}

@article{gomadam2007modeling,
  title={Modeling Li/ CF x-SVO Hybrid-Cathode Batteries},
  author={Gomadam, Parthasarathy M and Merritt, Donald R and Scott, Erik R and Schmidt, Craig L and Skarstad, Paul M and Weidner, John W},
  journal={Journal of the Electrochemical Society},
  volume={154},
  number={11},
  pages={A1058},
  year={2007},
  publisher={IOP Publishing}
}

@article{leising1994solid,
  title={Solid-state characterization of reduced silver vanadium oxide from the Li/SVO discharge reaction},
  author={Leising, Randolph A and Thiebolt III, William C and Takeuchi, Esther Sans},
  journal={Inorganic Chemistry},
  volume={33},
  number={25},
  pages={5733--5740},
  year={1994},
  publisher={ACS Publications}
}

@article{crespi1995characterization,
  title={Characterization of silver vanadium oxide cathode material by high-resolution electron microscopy},
  author={Crespi, AM and Skarstad, PM and Zandbergen, HW},
  journal={Journal of power sources},
  volume={54},
  number={1},
  pages={68--71},
  year={1995},
  publisher={Elsevier}
}

@article{ramasamy2006discharge,
  title={Discharge characteristics of silver vanadium oxide cathodes},
  author={Ramasamy, RP and Feger, C and Strange, T and Popov, BN},
  journal={Journal of applied electrochemistry},
  volume={36},
  pages={487--497},
  year={2006},
  publisher={Springer}
}

@article{van2013understanding,
  title={Understanding Li diffusion in Li-intercalation compounds},
  author={Van der Ven, Anton and Bhattacharya, Jishnu and Belak, Anna A},
  journal={Accounts of chemical research},
  volume={46},
  number={5},
  pages={1216--1225},
  year={2013},
  publisher={ACS Publications}
}

@article{srinivasan2004discharge,
  title={Discharge model for the lithium iron-phosphate electrode},
  author={Srinivasan, Venkat and Newman, John},
  journal={Journal of the Electrochemical Society},
  volume={151},
  number={10},
  pages={A1517},
  year={2004},
  publisher={IOP Publishing}
}

@article{hess2013shrinking,
  title={Shrinking annuli mechanism and stage-dependent rate capability of thin-layer graphite electrodes for lithium-ion batteries},
  author={He{\ss}, Michael and Nov{\'a}k, Petr},
  journal={Electrochimica Acta},
  volume={106},
  pages={149--158},
  year={2013},
  publisher={Elsevier}
}

@article{cogswell2012coherency,
  title={Coherency strain and the kinetics of phase separation in LiFePO4 nanoparticles},
  author={Cogswell, Daniel A and Bazant, Martin Z},
  journal={ACS nano},
  volume={6},
  number={3},
  pages={2215--2225},
  year={2012},
  publisher={ACS Publications}
}

@article{cogswell2013theory,
  title={Theory of coherent nucleation in phase-separating nanoparticles},
  author={Cogswell, Daniel A and Bazant, Martin Z},
  journal={Nano letters},
  volume={13},
  number={7},
  pages={3036--3041},
  year={2013},
  publisher={ACS Publications}
}

@article{cahn1958free,
  title={Free energy of a nonuniform system. I. Interfacial free energy},
  author={Cahn, John W and Hilliard, John E},
  journal={The Journal of chemical physics},
  volume={28},
  number={2},
  pages={258--267},
  year={1958},
  publisher={American Institute of Physics}
}

@article{cahn1961spinodal,
  title={On spinodal decomposition},
  author={Cahn, John W},
  journal={Acta metallurgica},
  volume={9},
  number={9},
  pages={795--801},
  year={1961},
  publisher={Elsevier}
}

@book{newman2021electrochemical,
  title={Electrochemical systems},
  author={Newman, John and Balsara, Nitash P},
  year={2021},
  publisher={John Wiley \& Sons}
}

@article{bard2001fundamentals,
  title={Fundamentals and applications},
  author={Bard, Allen J and Faulkner, Larry R and others},
  journal={Electrochemical methods},
  volume={2},
  number={482},
  pages={580--632},
  year={2001},
  publisher={Wiley New York}
}

@article{dreyer2016new,
  title={A new perspective on the electron transfer: recovering the Butler--Volmer equation in non-equilibrium thermodynamics},
  author={Dreyer, Wolfgang and Guhlke, Clemens and M{\"u}ller, R{\"u}diger},
  journal={Physical Chemistry Chemical Physics},
  volume={18},
  number={36},
  pages={24966--24983},
  year={2016},
  publisher={Royal Society of Chemistry}
}

@article{heubner2015investigation,
  title={Investigation of charge transfer kinetics of Li-Intercalation in LiFePO4},
  author={Heubner, C and Schneider, M and Michaelis, A},
  journal={Journal of Power Sources},
  volume={288},
  pages={115--120},
  year={2015},
  publisher={Elsevier}
}

@book{thomas2002mathematical,
  title={Mathematical modeling of lithium batteries},
  author={Thomas, Karen E and Newman, John and Darling, Robert M},
  year={2002},
  publisher={Springer}
}

@article{zhao2019population,
  title={Population dynamics of driven autocatalytic reactive mixtures},
  author={Zhao, Hongbo and Bazant, Martin Z},
  journal={Physical Review E},
  volume={100},
  number={1},
  pages={012144},
  year={2019},
  publisher={APS}
}

@article{son2015silicon,
  title={Silicon carbide-free graphene growth on silicon for lithium-ion battery with high volumetric energy density},
  author={Son, In Hyuk and Hwan Park, Jong and Kwon, Soonchul and Park, Seongyong and R{\"u}mmeli, Mark H and Bachmatiuk, Alicja and Song, Hyun Jae and Ku, Junhwan and Choi, Jang Wook and Choi, Jae-man and others},
  journal={Nature communications},
  volume={6},
  number={1},
  pages={7393},
  year={2015},
  publisher={Nature Publishing Group UK London}
}

@article{liu2014pomegranate,
  title={A pomegranate-inspired nanoscale design for large-volume-change lithium battery anodes},
  author={Liu, Nian and Lu, Zhenda and Zhao, Jie and McDowell, Matthew T and Lee, Hyun-Wook and Zhao, Wenting and Cui, Yi},
  journal={Nature nanotechnology},
  volume={9},
  number={3},
  pages={187--192},
  year={2014},
  publisher={Nature Publishing Group UK London}
}

@article{hu2008superior,
  title={Superior storage performance of a Si@ SiOx/C nanocomposite as anode material for lithium-ion batteries},
  author={Hu, Yong-Sheng and Demir-Cakan, Rezan and Titirici, Maria-Magdalena and M{\"u}ller, Jens-Oliver and Schl{\"o}gl, Robert and Antonietti, Markus and Maier, Joachim},
  journal={Angewandte Chemie International Edition},
  volume={47},
  number={9},
  pages={1645--1649},
  year={2008},
  publisher={Wiley Online Library}
}

@article{gu2012situ,
  title={In situ TEM study of lithiation behavior of silicon nanoparticles attached to and embedded in a carbon matrix},
  author={Gu, Meng and Li, Ying and Li, Xiaolin and Hu, Shenyang and Zhang, Xiangwu and Xu, Wu and Thevuthasan, Suntharampillai and Baer, Donald R and Zhang, Ji-Guang and Liu, Jun and others},
  journal={Acs Nano},
  volume={6},
  number={9},
  pages={8439--8447},
  year={2012},
  publisher={ACS Publications}
}

@article{kganyago2003structural,
  title={Structural and electronic properties of lithium intercalated graphite LiC 6},
  author={Kganyago, KR and Ngoepe, PE},
  journal={Physical Review B},
  volume={68},
  number={20},
  pages={205111},
  year={2003},
  publisher={APS}
}

@article{ohzuku1993formation,
  title={Formation of lithium-graphite intercalation compounds in nonaqueous electrolytes and their application as a negative electrode for a lithium ion (shuttlecock) cell},
  author={Ohzuku, Tsutomu and Iwakoshi, Yasunobu and Sawai, Keijiro},
  journal={Journal of The Electrochemical Society},
  volume={140},
  number={9},
  pages={2490},
  year={1993},
  publisher={IOP Publishing}
}

@article{harris2010direct,
  title={Direct in situ measurements of Li transport in Li-ion battery negative electrodes},
  author={Harris, Stephen J and Timmons, Adam and Baker, Daniel R and Monroe, Charles},
  journal={Chemical Physics Letters},
  volume={485},
  number={4-6},
  pages={265--274},
  year={2010},
  publisher={Elsevier}
}

@article{chen2014phase,
  title={A phase-field model coupled with large elasto-plastic deformation: application to lithiated silicon electrodes},
  author={Chen, Lei and Fan, Feifei and Hong, Liang and Chen, James and Ji, Yanzhou Z and Zhang, SL and Zhu, T and Chen, LQ},
  journal={Journal of The Electrochemical Society},
  volume={161},
  number={11},
  pages={F3164},
  year={2014},
  publisher={IOP Publishing}
}

@article{lu2016voltage,
  title={Voltage hysteresis of lithium ion batteries caused by mechanical stress},
  author={Lu, Bo and Song, Yicheng and Zhang, Qinglin and Pan, Jie and Cheng, Yang-Tse and Zhang, Junqian},
  journal={Physical Chemistry Chemical Physics},
  volume={18},
  number={6},
  pages={4721--4727},
  year={2016},
  publisher={Royal Society of Chemistry}
}

@article{verbrugge2015formulation,
  title={Formulation for the treatment of multiple electrochemical reactions and associated speciation for the lithium-silicon electrode},
  author={Verbrugge, Mark and Baker, Daniel and Xiao, Xingcheng},
  journal={Journal of The Electrochemical Society},
  volume={163},
  number={2},
  pages={A262},
  year={2015},
  publisher={IOP Publishing}
}

@article{wang2016investigation,
  title={Investigation of the chemo-mechanical coupling in lithiation/delithiation of amorphous Si through simulations of Si thin films and Si nanospheres},
  author={Wang, Miao and Xiao, Xinran},
  journal={Journal of Power Sources},
  volume={326},
  pages={365--376},
  year={2016},
  publisher={Elsevier}
}

@article{jo2010si,
  title={Si--graphite composites as anode materials for lithium secondary batteries},
  author={Jo, Yong Nam and Kim, Yeri and Kim, Jeom Soo and Song, Jun Ho and Kim, Ki Jae and Kwag, Chong Yun and Lee, Dong Jun and Park, Chul Wan and Kim, Young Jun},
  journal={Journal of Power Sources},
  volume={195},
  number={18},
  pages={6031--6036},
  year={2010},
  publisher={Elsevier}
}

@article{muller2018quantification,
  title={Quantification and modeling of mechanical degradation in lithium-ion batteries based on nanoscale imaging},
  author={M{\"u}ller, Simon and Pietsch, Patrick and Brandt, Ben-Elias and Baade, Paul and De Andrade, Vincent and De Carlo, Francesco and Wood, Vanessa},
  journal={Nature communications},
  volume={9},
  number={1},
  pages={2340},
  year={2018},
  publisher={Nature Publishing Group UK London}
}

@article{ai2020electrochemical,
  title={Electrochemical thermal-mechanical modelling of stress inhomogeneity in lithium-ion pouch cells},
  author={Ai, Weilong and Kraft, Ludwig and Sturm, Johannes and Jossen, Andreas and Wu, Billy},
  journal={Journal of The Electrochemical Society},
  volume={167},
  number={1},
  pages={013512},
  year={2020},
  publisher={The Electrochemical Society}
}

@inproceedings{popp2019benchmark,
  title={Benchmark, ageing and Ante-Mortem of SOTA CCylindrical lithium-ion cells},
  author={Popp, HARTMUT and GLANZ, GREGOR and HAMID, RAAD and ZHANG, NINGXIN},
  booktitle={14th international A3PS conference on eco-mobility},
  year={2019}
}

@article{liu2020multiphysics,
  title={Multiphysics coupled computational model for commercialized Si/graphite composite anode},
  author={Liu, Binghe and Jia, Yikai and Li, Jiani and Jiang, Hanqing and Yin, Sha and Xu, Jun},
  journal={Journal of Power Sources},
  volume={450},
  pages={227667},
  year={2020},
  publisher={Elsevier}
}

@article{lory2020probing,
  title={Probing silicon lithiation in silicon-carbon blended anodes with a multi-scale porous electrode model},
  author={Lory, PF and Mathieu, B and Genies, S and Reynier, Y and Boulineau, A and Hong, W and Chandesris, M},
  journal={Journal of The Electrochemical Society},
  volume={167},
  number={12},
  pages={120506},
  year={2020},
  publisher={IOP Publishing}
}

@article{popp2020ante,
  title={Ante-mortem analysis, electrical, thermal, and ageing testing of state-of-the-art cylindrical lithium-ion cells.},
  author={Popp, Hartmut and Zhang, Ningxin and Jahn, Marcus and Arrinda, Mikel and Ritz, Simon and Faber, Matthias and Sauer, Dirk Uwe and Azais, Philippe and Cendoya, Iosu},
  journal={Elektrotech. Informationstechnik},
  volume={137},
  number={4},
  pages={169--176},
  year={2020}
}

@article{crespi2001modeling,
  title={Modeling and characterization of the resistance of lithium/SVO batteries for implantable cardioverter defibrillators},
  author={Crespi, Ann and Schmidt, Craig and Norton, John and Chen, Kaimin and Skarstad, Paul},
  journal={Journal of the Electrochemical Society},
  volume={148},
  number={1},
  pages={A30},
  year={2001},
  publisher={IOP Publishing}
}

@article{leifer2007nuclear,
  title={Nuclear magnetic resonance and X-ray absorption spectroscopic studies of lithium insertion in silver vanadium oxide cathodes},
  author={Leifer, ND and Colon, A and Martocci, K and Greenbaum, SG and Alamgir, FM and Reddy, TB and Gleason, NR and Leising, RA and Takeuchi, ES},
  journal={Journal of the Electrochemical Society},
  volume={154},
  number={6},
  pages={A500},
  year={2007},
  publisher={IOP Publishing}
}

@article{garcia1994lithium,
  title={Lithium intercalation in Ag2V4O11},
  author={Garcia-Alvarado, F and Tarascon, JM},
  journal={Solid State Ionics},
  volume={73},
  number={3-4},
  pages={247--254},
  year={1994},
  publisher={Elsevier}
}

@article{grisolia2011density,
  title={Density functional theory investigations of the structural and electronic properties of Ag 2 V 4 O 11},
  author={Grisolia, Maricarmen and Rozier, Patrick and Benoit, Magali},
  journal={Physical Review B},
  volume={83},
  number={16},
  pages={165111},
  year={2011},
  publisher={APS}
}

@article{sauvage2010structural,
  title={Structural and transport evolution in the LixAg2V4O11 system},
  author={Sauvage, F and Bodenez, V and Vezin, Herve and Morcrette, M and Tarascon, J-M and Poeppelmeier, KR},
  journal={Journal of Power Sources},
  volume={195},
  number={4},
  pages={1195--1201},
  year={2010},
  publisher={Elsevier}
}

@article{takeuchi2001silver,
  title={Silver vanadium oxides and related battery applications},
  author={Takeuchi, Kenneth J and Marschilok, Amy C and Davis, Steven M and Leising, Randolph A and Takeuchi, Esther S},
  journal={Coordination Chemistry Reviews},
  volume={219},
  pages={283--310},
  year={2001},
  publisher={Elsevier}
}

@article{onoda2001crystal,
  title={Crystal structure and electronic properties of the Ag2V4O11 insertion electrode},
  author={Onoda, Masashige and Kanbe, Keisuke},
  journal={Journal of Physics: Condensed Matter},
  volume={13},
  number={31},
  pages={6675},
  year={2001},
  publisher={IOP Publishing}
}

@article{untereker2016power,
  title={Power sources and capacitors for pacemakers and implantable cardioverter-defibrillator},
  author={Untereker, Darrel F and Schmidt, CL and Jain, G and Tamirisa, PA and Hossick-Schott, J and Viste, M},
  journal={Clinical Cardiac Pacing, Defibrillation and Resynchronization Therapy, 5th ed.; Ellenbogen, KA, Wilkoff, BL, Kay, GN, Lau, C.-P., Auricchio, A., Eds},
  pages={251--269},
  year={2016}
}

@article{west1995lithium,
  title={Lithium insertion into silver vanadium oxide, Ag2V4O11},
  author={West, K and Crespi, AM},
  journal={Journal of power sources},
  volume={54},
  number={2},
  pages={334--337},
  year={1995},
  publisher={Elsevier}
}

@article{tiedemann1974electrochemical,
  title={Electrochemical behavior of the fluorographite electrode in nonaqueous media},
  author={Tiedemann, William},
  journal={Journal of The Electrochemical Society},
  volume={121},
  number={10},
  pages={1308},
  year={1974},
  publisher={IOP Publishing}
}

@article{davis2007simulation,
  title={Simulation of the Li--CF x System},
  author={Davis, Steven and Takeuchi, Esther S and Tiedemann, William and Newman, John},
  journal={Journal of the Electrochemical Society},
  volume={154},
  number={5},
  pages={A477},
  year={2007},
  publisher={IOP Publishing}
}

@article{farkhondeh2014mesoscopic,
  title={Mesoscopic modeling of Li insertion in phase-separating electrode materials: application to lithium iron phosphate},
  author={Farkhondeh, Mohammad and Pritzker, Mark and Fowler, Michael and Safari, Mohammadhosein and Delacourt, Charles},
  journal={Physical Chemistry Chemical Physics},
  volume={16},
  number={41},
  pages={22555--22565},
  year={2014},
  publisher={Royal Society of Chemistry}
}

@article{sasaki2013memory,
  title={Memory effect in a lithium-ion battery},
  author={Sasaki, Tsuyoshi and Ukyo, Yoshio and Nov{\'a}k, Petr},
  journal={Nature materials},
  volume={12},
  number={6},
  pages={569--575},
  year={2013},
  publisher={Nature Publishing Group UK London}
}

@article{takahashi2001characterization,
  title={Characterization of LiFePO4 as the cathode material for rechargeable lithium batteries},
  author={Takahashi, Masaya and Tobishima, Shinichi and Takei, Koji and Sakurai, Yoji},
  journal={Journal of Power Sources},
  volume={97},
  pages={508--511},
  year={2001},
  publisher={Elsevier}
}

@article{jacobsen1989li,
  title={Li insertion in CuyTiS2 spinels},
  author={Jacobsen, T and Zachau-Christiansen, B and West, K and Atlung, S},
  journal={Electrochimica acta},
  volume={34},
  number={10},
  pages={1473--1477},
  year={1989},
  publisher={Elsevier}
}

@article{mckinnon1984salting,
  title={Salting-out in intercalation compounds: removing copper from Cu3Mo6S8 by intercalating lithium},
  author={McKinnon, WR and Dahn, JR},
  journal={Solid state communications},
  volume={52},
  number={3},
  pages={245--248},
  year={1984},
  publisher={Elsevier}
}

@article{zandbergen1994two,
  title={Two structures of Ag2-xV4O11, determined by high resolution electron microscopy},
  author={Zandbergen, HW and Crespi, AM and Skarstad, PM and Vente, JF},
  journal={Journal of Solid State Chemistry},
  volume={110},
  number={1},
  pages={167--175},
  year={1994},
  publisher={Elsevier}
}

@article{allen2008analysis,
  title={Analysis of the FePO 4 to LiFePO 4 phase transition},
  author={Allen, JL and Jow, TR and Wolfenstine, J},
  journal={Journal of Solid State Electrochemistry},
  volume={12},
  pages={1031--1033},
  year={2008},
  publisher={Springer}
}

@article{allen2007kinetic,
  title={Kinetic study of the electrochemical FePO4 to LiFePO4 phase transition},
  author={Allen, Jan L and Jow, T Richard and Wolfenstine, Jeffrey},
  journal={Chemistry of materials},
  volume={19},
  number={8},
  pages={2108--2111},
  year={2007},
  publisher={ACS Publications}
}

@article{oyama2012kinetics,
  title={Kinetics of nucleation and growth in two-phase electrochemical reaction of Li x FePO4},
  author={Oyama, Gosuke and Yamada, Yuki and Natsui, Ryu-ichi and Nishimura, Shin-ichi and Yamada, Atsuo},
  journal={The Journal of Physical Chemistry C},
  volume={116},
  number={13},
  pages={7306--7311},
  year={2012},
  publisher={ACS Publications}
}

@article{rowlinson1979translation,
  title={Translation of JD van der Waals'“The thermodynamik theory of capillarity under the hypothesis of a continuous variation of density”},
  author={Rowlinson, John S},
  journal={Journal of Statistical Physics},
  volume={20},
  pages={197--200},
  year={1979},
  publisher={Springer}
}

@article{zhang2015progress,
  title={Progress towards high-power Li/CF x batteries: electrode architectures using carbon nanotubes with CF x},
  author={Zhang, Qing and Takeuchi, Kenneth J and Takeuchi, Esther S and Marschilok, Amy C},
  journal={Physical Chemistry Chemical Physics},
  volume={17},
  number={35},
  pages={22504--22518},
  year={2015},
  publisher={Royal Society of Chemistry}
}

@article{liu2019brief,
  title={A brief review for fluorinated carbon: synthesis, properties and applications},
  author={Liu, Yifan and Jiang, Lingyan and Wang, Haonan and Wang, Hong and Jiao, Wei and Chen, Guozhang and Zhang, Pinliang and Hui, David and Jian, Xian},
  journal={Nanotechnology Reviews},
  volume={8},
  number={1},
  pages={573--586},
  year={2019},
  publisher={De Gruyter}
}

@article{john1961spinodal,
  title={On spinodal decomposition Sur la decomposition spinodale {\"U}ber die umsetzung an der spinodalen},
  author={John, WC},
  journal={Acta Metall},
  volume={9},
  pages={795--801},
  year={1961}
}

@article{mond2014cardiac,
  title={The cardiac implantable electronic device power source: evolution and revolution},
  author={Mond, Harry G and Freitag, Gary},
  journal={Pacing and Clinical Electrophysiology},
  volume={37},
  number={12},
  pages={1728--1745},
  year={2014},
  publisher={Wiley Online Library}
}

@article{smith2017intercalation,
  title={Intercalation kinetics in multiphase-layered materials},
  author={Smith, Raymond B and Khoo, Edwin and Bazant, Martin Z},
  journal={The Journal of Physical Chemistry C},
  volume={121},
  number={23},
  pages={12505--12523},
  year={2017},
  publisher={ACS Publications}
}

@article{delmas2008lithium,
  title={Lithium deintercalation in LiFePO4 nanoparticles via a domino-cascade model},
  author={Delmas, Claude and Maccario, Magalie and Croguennec, Laurence and Le Cras, Fr{\'e}d{\'e}ric and Weill, Fran{\c{c}}ois},
  journal={Nature materials},
  volume={7},
  number={8},
  pages={665--671},
  year={2008},
  publisher={Nature Publishing Group UK London}
}

@article{brunetti2011confirmation,
  title={Confirmation of the domino-cascade model by LiFePO4/FePO4 precession electron diffraction},
  author={Brunetti, G and Robert, D and Bayle-Guillemaud, P and Rouviere, JL and Rauch, EF and Martin, JF and Colin, JF and Bertin, F and Cayron, C},
  journal={Chemistry of Materials},
  volume={23},
  number={20},
  pages={4515--4524},
  year={2011},
  publisher={ACS Publications}
}

@article{robert2013multiscale,
  title={Multiscale phase mapping of LiFePO4-based electrodes by transmission electron microscopy and electron forward scattering diffraction},
  author={Robert, Donatien and Douillard, Thierry and Boulineau, Adrien and Brunetti, Guillaume and Nowakowski, Pawel and Venet, Denis and Bayle-Guillemaud, Pascale and Cayron, Cyril},
  journal={ACS nano},
  volume={7},
  number={12},
  pages={10887--10894},
  year={2013},
  publisher={ACS Publications}
}

@article{dreyer2010thermodynamic,
  title={The thermodynamic origin of hysteresis in insertion batteries},
  author={Dreyer, Wolfgang and Jamnik, Janko and Guhlke, Clemens and Huth, Robert and Mo{\v{s}}kon, Jo{\v{z}}e and Gaber{\v{s}}{\v{c}}ek, Miran},
  journal={Nature materials},
  volume={9},
  number={5},
  pages={448--453},
  year={2010},
  publisher={Nature Publishing Group UK London}
}

@article{dreyer2011hysteresis,
  title={Hysteresis and phase transition in many-particle storage systems},
  author={Dreyer, Wolfgang and Guhlke, Clemens and Herrmann, Michael},
  journal={Continuum Mechanics and Thermodynamics},
  volume={23},
  pages={211--231},
  year={2011},
  publisher={Springer}
}

@article{dreyer1982study,
  title={A study of equilibria of interconnected balloons},
  author={Dreyer, W and M{\"u}ller, I and Strehlow, P},
  journal={The Quarterly Journal of Mechanics and Applied Mathematics},
  volume={35},
  number={3},
  pages={419--440},
  year={1982},
  publisher={Oxford University Press}
}

\end{document}